\documentclass[journal=jpcbfk,manuscript=article]{achemso}
\usepackage{longtable}
\usepackage{multirow}
\usepackage{amsmath}
\usepackage[usenames,dvipsnames]{color}
\usepackage{soul}
\usepackage{threeparttable}
\usepackage{graphicx}
\usepackage{amsmath,amssymb}
\usepackage{caption}
\usepackage{color}
\usepackage{xcolor}
\usepackage{dcolumn}
\usepackage{bm}
\usepackage{float}
\usepackage{subcaption}
\usepackage{textcomp}
\usepackage{url}
\usepackage[normalem]{ulem}

\usepackage{microtype}

\usepackage[unicode=true,
            bookmarks=true,
            bookmarksnumbered=true,
            bookmarksopen=true,
            bookmarksopenlevel=2,
            breaklinks=false,
            pdfborder={0 0 1},
            backref=false,
            colorlinks=true,
            hidelinks
            ]{hyperref}

\makeatletter

\author{Kai T\"opfer} \affiliation[University of Basel]{Department of
  Chemistry, University of Basel, Klingelbergstrasse 80, CH-4056
  Basel, Switzerland.}

\author{Eric Boittier} \affiliation[University of Basel]{Department of
  Chemistry, University of Basel, Klingelbergstrasse 80, CH-4056
  Basel, Switzerland.}

\author{Mike Devereux} \affiliation[University of Basel]{Department
  of Chemistry, University of Basel, Klingelbergstrasse 80, CH-4056
  Basel, Switzerland.}

\author{Andrea Pasti}  \affiliation[University of
  Zurich]{Department of Chemistry, University of Z\"urich, CH-8000
  Z\"urich, Switzerland.}

\author{Peter Hamm}  \affiliation[University of
  Zurich]{Department of Chemistry, University of Z\"urich, CH-8000
  Z\"urich, Switzerland.}

\author{Markus Meuwly} \affiliation[University of Basel]{Department of
  Chemistry, University of Basel, Klingelbergstrasse 80, CH-4056
  Basel, Switzerland.}
\email{m.meuwly@unibas.ch}

\title{Force Fields for Deep Eutectic Mixtures: Application to
  Structure, Thermodynamics and 2D-Infrared Spectroscopy}

\date{\today}

\begin{document}
\date{\today}

\begin{abstract}
Parametrizing energy functions for ionic systems can be
challenging. Here, the total energy function for an eutectic system
consisting of water, SCN$^-$, K$^+$ and acetamide is improved
vis-a-vis experimentally measured properties. Given the importance of
electrostatic interactions, two different types of models are
considered: the first (model {\bf M0}) uses atom-centered multipole
whereas the other two (models {\bf M1} and {\bf M2}) are based on
fluctuating minimal distributed charges (fMDCM) that respond to
geometrical changes of SCN$^-$. The Lennard-Jones parameters of the
anion are adjusted to best reproduce experimentally known hydration
free energies and densities which are matched to within a few percent
for the final models irrespective of the electrostatic
model. Molecular dynamics simulations of the eutectic mixtures with
varying water content (between 0 and 100 \%) yield radial distribution
functions and frequency correlation functions for the CN-stretch
vibration. Comparison with experiments indicate that models based on
fMDCM are considerably more consistent that those using
multipoles. Computed viscosities from models {\bf M1} and {\bf M2} are
within 30 \% of measured values and their change with increasing water
content is consistent with experiments. This is not the case for model
{\bf M0}.
\end{abstract}

\section{Introduction}
Characterizing the energetics, structural dynamics and spectroscopy of
heterogeneous systems in the condensed phase is challenging from a
computational and an experimental perspective. Typical examples of
such endeavours are hydrated proteins in a representative environment
including the necessary anions and cations, ionic liquids, or eutectic
mixtures. From a computational perspective the main challenges are the
accuracy of the energy function that is required to carry out
atomistic simulations and the time scale on which such simulations can
be run. On the other hand, from an experimental viewpoint, one of the
greatest challenges is the fact that usually the ``full system'' is
probed which may lead to spectral congestion if optical spectroscopy
is used as an example. Also, covering multiple time scales from the
femtosecond to the second time scale is demanding, although recent
progress has been made here.\cite{hamm:2023}\\

\noindent
For meaningful atomistic simulations sufficiently accurate energy
functions are a prerequisite. Although there has been impressive
progress in directly solving the electronic Schr\"odinger equation for
increasingly large systems, doing so along a molecular dynamics
simulation is usually unfeasible beyond the multiple-ten-picosecond
time scale, in particular for methods superior to density functional
theory. On the other hand, empirical energy functions (``force
  fields'') are computationally advantageous to evaluate and are being
  used in molecular dynamics (MD) simulations which solve the
  Newtonian equations of motions. MD simulations have been used to
investigate processes ranging from protein folding,\cite{shaw:2014}
ligand binding,\cite{simonson:2002} crowding in cellular
environments\cite{feig:2019} to characterizing spectroscopic
properties of solutes and peptides and reactions in the gas phase and
in solution.\cite{MM.jcp:2020} However one of the challenges remains
to develop energy functions that retain the precision of the QM
methods they are often based on.\\

\noindent
One- or multi-dimensional vibrational spectroscopy is a powerful means
to characterize the structural dynamics of complex
systems.\cite{pines:2005,2dir:2011} The molecular-level understanding
and interpretation of the spectral signatures for such systems is
greatly aided by a combined simulation/experiment
approach.\cite{MM.cn:2013,MM.eutectic:2022} The atomistic simulations
often rely on energy functions that parametrize the bonded and
nonbonded interactions. Traditionally, empirical energy functions use
harmonic springs for chemical bonds and valence angles, periodic
functions for dihedrals, an atom-centered point charge-based model for
charges and a Lennard-Jones representation for van der Waals
interactions together with additional, more purpose-tailored
terms.\cite{mackerell2004} For applications to vibrational
  spectroscopy, going beyond the harmonic approximation to describe
  chemical bonds is necessary to include effects of mechanical
  anharmonicity. Such improvements can, e.g., be achieved through the
  use of machine learning-based approaches, in particular if
reference data from high-level electronic structure calculations are
available.\cite{MM.n3:2019,MM.jcp:2020,MM.rkhs:2020,nandi:2019,li:2014}
For the nonbonded interactions - which include electrostatic and van
der Waals terms - more physics-based models that go beyond the
standard representations have been developed.\\

\noindent
The first-order treatment of the electrostatic interaction is based on
atom-centered point charges for Coulomb interactions. Such pair
interactions can be rapidly computed but lack the accuracy for
describing anisotropic contributions to the charge
density.\cite{Stone2013} Including higher-order atomic multipoles
improves the accuracy but at the expense of increased computational
cost and implementation
complexity.\cite{Handley2009,MM.mtp:2013,Devereux2014,bereau:2016}
Accounting for polarizability is another contribution that has been
recently included in empirical force fields and shows much promise for
further improvements of the computational models.\cite{ren:2019} From
an empirical force field perspective the van der Waals interactions
are often represented as Lennard-Jones terms with {\it ad hoc}
(Lorentz-Berthelot) combination rules. Alternative and potentially
improved representations are the buffered 14-7
parametrization\cite{halgren:1992} and modified combination
rules\cite{mason:1988,millie:2001}\\

\noindent
Deep eutectic solvents (DESs) are multicomponent mixtures
  comprising hydrogen bond acceptors (HBAs) and hydrogen bond donors
  (HBDs) at particular molar
  ratios.\cite{abbott2003DES,marcus2019trends,martins2019defdes} In
  general, DESs are characterized by a pronounced depression of their
  melting point compared to their components and remain in the liquid
  phase over a wider temperature range.\cite{smith:2014} If the
  mixtures contain ions, the intermolecular interactions involve
  pronounced electrostatic contributions which is also - in part - due
  to crowding. The particular mixture considered here consists of
  water, acetamide and KSCN which is present as solvated K$^+$ and
  SCN$^-$ (thiocyanate) ions.\cite{isaac:1988,kalita:1998} Acetamide
  forms low-temperature eutectics with a wide range of inorganic salts
  and the resulting non-aqueous solvents have a high ionicity. Such
  mixtures have also been recognized as excellent solvents and molten
  acetamide is known to dissolve inorganic and organic compounds. The
  SCN$^-$ anion is an ideal spectroscopic probe because the CN-stretch
  vibration absorbs in an otherwise empty region of the infrared
  region. Previous work has taken advantage of these spectral features to
  probe the
  effect of water addition to urea/choline chloride and in
  acetamide/water mixtures.\cite{sakpal:2021,MM.eutectic:2022}\\

\noindent
Traditionally, energy functions for DES have been optimized by
  starting from a conventional energy function and readjusting
  specifically partial charges and in
  some cases van der Waals parameters.\cite{ferreira:2016,padua:2022,
    maglia:2021,doherty:2018,jeong:2021,garcia:2015,zhang:2022,velez:2022}
The present work aims at parametrizing atomistic force fields using
state-of-the art methods by combining machine learning-based
approaches for bonded and nonbonded terms, refinement of the
Lenard-Jones interactions with respect to thermodynamic data and
validation on structural, spectroscopic and thermodynamic
measurements. First, the methods are presented, followed by the
reparametrization and validation of the energy functions. Next,
extended MD simulations are analyzed with respect to pair distribution
functions and the frequency fluctuation correlation functions from
experiment and simulations are compared. This is followed by a
discussion and conclusions.\\

\section{Computational and Experimental Methods}

\subsection{Simulation Setup}
Molecular dynamics simulations were performed using the CHARMM
program.\cite{Charmm-Brooks-2009} The molar composition of the systems
was changed by modifying the number of water and acetamide molecules
while keeping constant the total concentration of K$^+$ and
SCN$^{-}$. The number of molecules and molar fractions are reported in
Table S1. Bonds involving hydrogen atoms were
constrained using the SHAKE algorithm.\cite{shake77} For the nonbonded
interactions the cutoff was 14~\AA\/ and electrostatic interactions
were treated using the Particle Mesh Ewald algorithm.\cite{Darden1993}
In total, 5 independent random initial configurations for each of the
9 system compositions were set up using PACKMOL.\cite{martinez:2009}
Following 100~ps of heating and equilibrium simulation, each,
production simulations in the \emph{NpT} ensemble at 300~K and normal
pressure (1 atm) were performed for 5\,ns with a time step of 1\,fs
using the leap-frog integrator and a Hoover 
thermostat within the extended system constant pressure and
temperature algorithm as implemented in CHARMM.\cite{Hoover1985,Brooks1995}
The mass of the pressure piston and piston collision frequency were
set to $406$\,u and 5\,ps$^{-1}$, respectively, and the mass of the
thermal piston to $4060$\,kcal/mol\,ps$^2$.
For each system composition a total of $25$\,ns were sampled.\\

\subsection{Inter- and Intramolecular Interactions}
The heterogeneous system considered in the present work (water,
acetamide, K$^+$, SCN$^{-}$) was described by a combination of
parametrizations based on the all-atom force field CHARMM36
(CGenFF)\cite{cgenff}, the corresponding TIP3P water
model\cite{TIP3P-Jorgensen-1983} to be used together with CGenFF, and
optimized parameters for the ions. Two sets of nonbonding
van-der-Waals (vdW) parameters for the potassium cation K$^+$ with an
assigned atom charge of $+1.0$ were adopted from previous
work\cite{bian:2013,lund:2018} and acetamide was described by
CGenFF. Dedicated parametrizations for bonding and nonbonding terms
were used for the thiocyanate anion (SCN$^-$) as described next.\\

\noindent
{\it Bonded Interactions:} The bonding potential for the thiocyanate
anion (SCN$^-$) was a reproducing kernel Hilbert space
(RKHS)\cite{rabitz:1996,MM.rkhs:2017} representation based on {\it ab
  initio} data at the PNO-LCCSD(T)-F12/aug-cc-pVTZ-F12 level of theory
using the MOLPRO program package.\cite{werner:2017,werner:2020} The
geometry of SCN$^-$ was described in terms of Jacobi coordinates
$(r,R,\theta)$ where $r$ is the CN separation, $R$ is the distance
between S and the center of mass of CN and $\theta$ is the angle
between the vectors $\vec{r}$ and $\vec{R}$. Reference energies were
determined on an equidistant grid defined by $r = [1.072,
  1.322]$\,\AA\/ (8 points), $R = [2.009, 2.609]$\,\AA\/ (9 points),
and $\theta = [150.0, 180.0]^\circ$\/ (6 points). This yields a total
of 432 grid points. Energies for grid points within $\sim
30$\,kcal/mol of the equilibrium structure
($r_\mathrm{eq}=1.172$\,\AA, $R_\mathrm{eq}=2.309$\,\AA\/ and
$\theta_\mathrm{eq}=180.0^\circ$) were retained for constructing the
RKHS. The root mean squared error (RMSE) between reference and the
RKHS representation was $0.182$\,kcal/mol. Two-dimensional cuts
through the RKHS representation for SCN$^-$ along Jacobi distances $r$
and $R$ for different angles $\theta$ are shown in Figure
S1.  \\

\noindent
{\it Electrostatic Interactions:} For the electrostatic interactions
the electrostatic potential (ESP) of SCN$^-$ was represented as a
flexible minimally distributed charge model
(fMDCM).\cite{MM.fmdcm:2022} For this, 8 point charges were
distributed around the SCN$^-$ atoms with positions within the local
axis frame determined by 3rd order polynomial functions $f(x)$ with
$x=1-\cos^2 \theta$ and the SCN$^-$ bond angle $\theta$. The 96
parameters of the 24 polynomial functions - 4 parameters per
polynomial for each Cartesian coordinate (3) and distributed charge
(8) - were optimized to best reproduce the reference ESPs for
different SCN$^-$ conformations. The reference ESPs on a range of
geometries were determined at the M06-2X/aug-cc-pVTZ level of theory
using the Gaussian suite of programs.\cite{gaussian16} The grid
contained 289 geometries, including 32 SCN valence angles ($\theta =
[60^\circ, 180^\circ$]) and 3 bond lengths for the SC ($r_\mathrm{SC}
= \{1.646, 1.661, 1.676\}$\,\AA) and CN ($r_\mathrm{CN} = \{1.182,
1.192, 1.202\}$\,\AA) separations.\\

\begin{figure}
  \centering
  \includegraphics[width=0.50\textwidth]{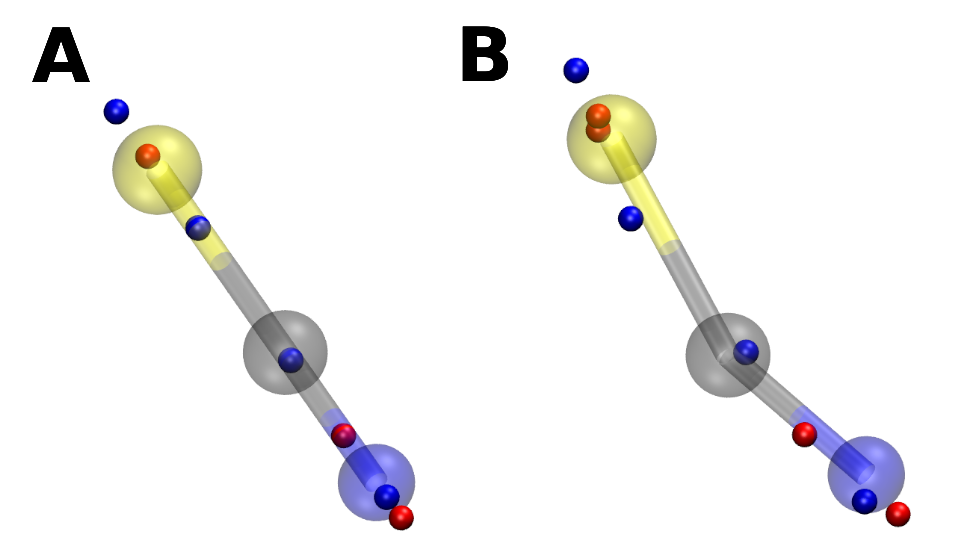}
\caption{Distributed charge positions (red spheres as negative charges
  and blue spheres positive charges) predicted by the fMDCM model to
  represent the ESP of SCN$^-$ (transparent spheres) for the (A)
  linear equilibrium conformation and (B) at a valence angle $\theta =
  160^\circ$.}
\label{fig_fmdcm}
\end{figure}

\noindent
Starting from the initial 8 charge MDCM model the reference ESP was
fit for SCN$^-$ in a slightly bent conformation ($\theta = 171^\circ$)
according to the optimization scheme presented in
Refs. \citenum{MM.mdcm:2017,MM.mdcm:2020}. Starting from this model as
an initial guess for the polynomial parameters, the quasi-Newton BFGS
minimization method\cite{Nocedal2006Numerical} was used to optimize
the polynomial parameters by minimizing the RMSE between reference and
computed ESP grid point values for each conformation. The total RMSE
was the sum of the RMSE per conformation $i$ weighted by a factor $w_i
= 1 + (V_i - V_\mathrm{min})^4$ to prioritize an accurate ESP fit for
low-energy conformations.  Here, $V_i$ was the {\it ab initio} energy
of SCN$^-$ in conformation $i$ relative to the energy of the
equilibrium conformation $V_\mathrm{min}$.  Furthermore, the 8
distributed charges were constrained to lie only along the bond axis
for linear SCN$^-$ conformations, the charge displacement $d$ from the
closest reference atom and its charge magnitude $c$ within $d \le
1.0$\,a$_0$ and $c = [-1.0, +1.0]e$ respectively. Figure
\ref{fig_fmdcm}A shows the the distributed charge position predicted
by the fMDCM model for SCN$^-$ in linear equilibrium conformation
where all charges a aligned along the SCN$^-$ axis.  For angled
SCN$^-$ conformation such as in Figure \ref{fig_fmdcm}B with a bond
angle of 160$^\circ$ the charge positions deviates from the bond
axes. Note the two negative distributed charges (red spheres) closest
to the sulfur atom (transparent yellow sphere) which are merged into
the same position for linear SCN$^-$, see Figure \ref{fig_fmdcm}A, and
become symmetrically lifted and mirrored at the SCN$^-$ plane in
Figure \ref{fig_fmdcm}B.\\

\noindent
The RMSE between the fMDCM model and the reference \emph{ab initio}
ESP was $2.558$\,kcal/mol on average and $1.350$\,kcal/mol for the
linear equilibrium structure of SCN$^-$. Following the
literature,\cite{petersen2005scnpol} the diagonal elements of computed
polarizability $\boldsymbol{\alpha}$ for SCN$^-$ in the gas phase
(M06-2X/aug-cc-pVTZ) were $\alpha_{xx}=\alpha_{K}=6.2$\,\AA$^3$\/ and
$\alpha_{zz}=10.9$\,\AA$^3$\/ to yield an average isotropic
polarizability of $\alpha_\mathrm{iso} = 7.8$\,\AA$^3$ However,
computational studies have shown a reduction in the polarizability for
anions from gas phase to water solution by between 5\% and 34\%,
respectively.\cite{salvador2003pol,jungwirth2002pol} Thus, a reduced
polarizability of $6$\,\AA$^3$\/ was equally distributed to the atoms
of SCN$^-$ with $2$\,\AA$^3$\/ each.\\

\noindent
In addition to the new fMDCM model, a multipolar
(MTP)\cite{MM.mtp:2013,mm.mtp2:2016,mm.mtp:2017} representation of
SCN$^-$ used in previous work was used as well.\cite{MM.eutectic:2022}
The electrostatic atomic moments up to the quadrupole moment were
optimized to fit the ESP of SCN$^-$ in its equilibrium conformation
computed at the MP2/aug-cc-pVDZ level of theory.\\

\noindent
{\it Van-der-Waals Interactions:} Changing the electrostatic model for
SCN$^{-}$ from point charges to a more elaborate fMDCM or MTP model
also requires to readjust the Lennard-Jones
parameters.\cite{MM.ff:2024} Three models were considered in the
following for SCN$^{-}$. Model ({\bf M0}) was based on MTP with LJ
parameters originally from Bian {\it et al.}\cite{bian:2013} and
scaled to best reproduce measured thermodynamic properties. Model {\bf
  M1} used fMDCM electrostatic instead of MTP and initial LJ
parameters from Tesei {\it et al.},\cite{lund:2018} and model {\bf M2}
was a combination of fMDCM and optimized LJ parameters with respect to
reference {\it ab initio} interaction energies as described
below. Experimentally observed density of aqueous KSCN solution and
the hydration free energy of SCN$^{-}$ in water were then determined
for each of the models {\bf M0}, {\bf M1} and {\bf M2}. For the K$^+$
cation the LJ parameters were those from the respective parameter set
in the literature (Ref. \citenum{bian:2013} for {\bf M0} and {\bf M2}
and Ref. \citenum{lund:2018} for {\bf M1}).\\

\noindent

\noindent
For model {\bf M2}, also based on the fMDCM representation for
  the SCN$^-$ anion, a different approach was taken. Here, the LJ
  parameters $\epsilon$ and $r_\mathrm{min}$ were adjusted to best
  match counterpoise-corrected\cite{boys:1970} interaction energies
  from electronic structure calculations. The systems considered
  included one SCN$^-$ anion surrounded by (i) 16 water molecules;
  (ii) 14 water molecules with one additional SCN$^-$, (iii)
  14 water molecules with one K$^+$ ion; (iv) 12 water molecules
  with both one additional SCN$^-$ and K$^+$ ion. For each type
  of system (i) to (iv) 50 independent conformations were extracted
  randomly from previous MD simulations.\cite{MM.eutectic:2022}
  Reference interaction energies were then determined at the
  M06-2X/aug-cc-pVTZ level of theory using Gaussian
  program.\cite{gaussian16} Higher levels of quantum chemical theory,
  such as coupled cluster energies, are not feasible due to the
  unfavourable scaling of the computations with system and basis set
  size.\\

\subsection{Analysis}
Hydration free energies $\Delta G_{\mathrm{hyd}}$ for the SCN$^-$
anion in water solvent were computed from thermodynamic
integration.\cite{MM.cn:2013} One SCN$^-$ anion was sampled in the gas
phase and in pure water. The condensed-phase simulations were carried
out in the \emph{NpT} ensemble with 997 water molecules (cubic box
size $\sim 30^3$\,\AA$^3$).\cite{straatsma:1988,MM.cn:2013}
Simulations with 24 different coupling parameters $\lambda \in (0, 1)$
for electrostatic and vdW interactions, respectively, were performed
for each a gas phase and solvated SCN$^-$. Initial conditions for
these simulation were taken from an unbiased simulation, equilibrated
for 50\,ps with the respective coupling parameter $\lambda$ and run
for another 150\,ps for statistical sampling.  In general, the
hydration free energy is accumulated from
\begin{equation}
     \Delta G_{\mathrm{hyd}} = \sum_\lambda [ (
       H_\mathrm{solv}^\mathrm{elec}(\lambda) -
       H_\mathrm{gas}^\mathrm{elec}(\lambda) ) + (
       H_\mathrm{solv}^\mathrm{vdW}(\lambda) -
       H_\mathrm{gas}^\mathrm{vdW}(\lambda) ) ] \Delta \lambda
\end{equation}
For a triatomic such as SCN$^-$,
$H_\mathrm{gas}^\mathrm{elec}(\lambda) =
H_\mathrm{gas}^\mathrm{vdW}(\lambda) = 0$ due to the 1-2 and 1-3
nonbonded interaction exclusion.\cite{mackerell:2004} Therefore, only
$H_\mathrm{solv}^\mathrm{elec}(\lambda)$ and
$H_\mathrm{solv}^\mathrm{vdW}(\lambda)$ needed to be accumulated.\\

\noindent
The density of an aqueous KSCN solution was determined from 500 ps
simulations in the $NpT$ ensemble. The system composition corresponds
to the 100\% one reported in Table S1 with a KSCN
molality of $b(\mathrm{KSCN})=3.821$\,mol/kg. The equilibrium
simulation box volume is computed as the average box volume within the
last 100\,ps of the simulations (400--500\,ps in the production run).
Convergence within the reported precision was checked by comparing
with the average taken from the results of the full production run of
500\,ps.\\

\noindent
For computing the frequency fluctuation correlation function (FFCF),
the frequency trajectories $\omega_i(t)$ for each oscillator $i$ are
required. These were determined from an instantaneous normal mode
(INM)\cite{stratt:1994} analysis for the CN vibrational frequencies
$\omega_i$ for all SCN$^-$ ions on snapshots of the trajectories every
$100$\,fs along the first 2\,ns of each production simulation (10\,ns
in total for each system composition).  For the INM analysis the
structure of every SCN$^-$ ion was optimized while freezing the
position of all remaining atoms in the system, followed by a normal
mode analysis using the same force field that was employed for the MD
simulations. Previously, such an approach has been validated for
N$_3^-$ in solution by comparing with rigorous quantum bound state
calculations.\cite{MM.n3:2019}\\

\noindent
From the frequency trajectory $\omega_i(t)$ for each oscillator the
FFCF $\delta \omega_i(t) = \omega_(t) - < \omega_i >$ was determined
which contains information on relaxation time scales corresponding to
the solvent dynamics around the solute. The FFCFs were fit to an
empirical expression\cite{kozinski:2007,MM.cn:2013}
\begin{equation}
  \langle \delta \omega(t) \delta \omega(0) \rangle = a_{1}
  e^{-t/\tau_{1}} + a_{2} e^{-t/\tau_{2}} + \Delta_0^2
\label{eq:ffcffit}
\end{equation}
using an
automated curve fitting function (\texttt{scipy.optimize.curve\_fit})
from the SciPy library using the default trust region reflective 
algorithm.\cite{2020SciPy-NMeth}
Here, $a_{i}$, $\tau_{i}$, $\gamma$ and
$\Delta_0^2$ are the amplitudes, decay time scales, phase and
asymptotic value of the FFCF.\\

\subsection{2D-IR Spectroscopy}
The experimental data shown in this paper (Figure \ref{fig_ffcf})
originate from the same raw data as those of
Ref.~\cite{MM.eutectic:2022}, this time extracting the FFCF from the
2D-IR spectra. Technical details of the 2D~IR spectrometer used
  to obtain the raw data are described in Ref.~\citenum{Helbing2011}. A
50\%/50\% mixture of two isotopologues of the salt, KS$^{12}$C$^{14}$N
and KS$^{13}$C$^{15}$N has been investigated. Solutions were
prepared with constant total salt concentration of 4.4~M, and the
molar ratio of D$_2$O vs acetamide varied from 0\% to 100\% in steps
of 10\%.\\

\noindent
Very generally speaking, the 2D~IR spectrum of two coupled modes consist of two diagonal contributions for each mode, and two cross peak contributions that appear when energy transfer occurs between both modes. Furthermore, each one of these contributions consist of a pair of a negative and positive peak, i.e., ground state bleach and excited state absorption, respectively. Hence, eight peaks exist in total.\cite{2dir:2011}
To extract the essential information from the 2D~IR spectra, i.e., the
FFCF, each one of these 2D~IR peaks was modelled as a correlated 2D-Gaussian
function:
\begin{align} \label{sieq:2DGC}
	G(\omega_{1},\omega_{3}) &= A \exp \biggl\{
        -\frac{1}{2(1-c^{2})}
        \biggl[\biggl(\frac{\omega_{1}-\omega_{1,0}}{\Delta{\omega}}\biggr)^{2}
          \\ +
          &\biggl(\frac{\omega_{3}-\omega_{3,0}}{\Delta\omega}\biggr)^{2}
          - \frac{
            2c(\omega_{1}-\omega_{1,0})(\omega_{3}-\omega_{3,0}) }{
            \Delta\omega^2} \biggr] \biggr\}, \nonumber
\end{align}
where $A$ is an amplitude, $\omega_{i,0}$ the center frequencies along frequency axis $i$, $\Delta\omega$ the frequency width, and $c$ a correlation coefficient.\cite{Guo2015} In the limiting case of $c=0$, the 2D Gaussian is uncorrelated and a circular peak would appear. For $c=1$ the 2D Gaussian is fully correlated with the peak stretched along the diagonal and widths $\Delta\omega$ in the diagonal direction and 0 in the anti-diagonal direction. To minimize the total number of parameters in the global fitting, many of the parameters for modeling the eight peaks as Gaussian functions were constrained to be equal.
That is, two pairs
of diagonal peaks were defined for the two isotopologues with
different central frequencies $\omega_{i,0}$ and amplitudes $A$, but
identical width $\Delta\omega$ and correlation $c$, since the latter
two parameters describe the identical interaction of the ions with the
solvent environment. For each diagonal contribution, the ground state
bleach was positioned on the diagonal (i.e., $\omega_{1,0}=\omega_{3,0}$), while the excited state absorption has been down-shifted in the
$\omega_3$-direction by the anharmonicity
$\Delta=27~\rm{cm}^{-1}$. The amplitudes $A$ and correlations $c$ of
the two contributions to each diagonal peak were assumed to be the
same. The center frequencies and widths of the cross peak pairs were assumed to be the same as for the corresponding diagonal peaks. The
amplitudes $A$ of the two cross peak pairs were set to be the same,
and $c=0$ was assumed for the correlation, since cross-relaxation
randomizes the frequency.\\

\noindent
The property of interest here is the frequency correlation $c(t_2)$ of
the diagonal peaks as a function of the waiting time $t_2$ between the
pump-pulse pair and the probe pulse. It evolves from ideally 1 at $t_2=1$ to 0 for $t_2\rightarrow\infty$, however, dynamics faster than few
100~fs is hidden in the homogeneous dephasing time $T_2$ and cannot be observed.\cite{2dir:2011}  Beyond a few
100~fs, $c(t_2)$ reflects the normalized FFCF and can be compared to results from
MD simulations after proper scaling.

\section{Results}
\subsection{Adaptation and Validation of the LJ Parameters}
Three models {\bf M0} to {\bf M2} with different treatment of the
electrostatics (MTP or fMDCM) and corresponding LJ parameters for
SCN$^-$ and K$^+$ were considered. Details on the models are given in
Tables S2 to S4. The
experimental reference data required for models {\bf M0} and {\bf M1}
were the density of aqueous KSCN solution
($\rho_\mathrm{exp}=1.139$\,g/cm$^3$) at the equivalent KSCN molality
of $b(\mathrm{KSCN})=3.821$\,mol/kg\cite{albright:1992} and the
hydration free energy $\Delta G_\mathrm{hyd}$. For SCN$^{-}$ an
estimated value of $\Delta G_\mathrm{hyd} \sim -72$ kcal/mol is
available\cite{marcus:1997} which compares with related anions such as
HS$^-$ ($-74.0$\,kcal/mol), N$_3^-$ ($-72.0$\,kcal/mol)
\cite{pearson:1986}, or CN$^-$ ($-72.0 \pm
0.7$\,kcal/mol)\cite{pliego:2000}.\\

\begin{figure}
  \centering
  \includegraphics[width=0.75\textwidth]{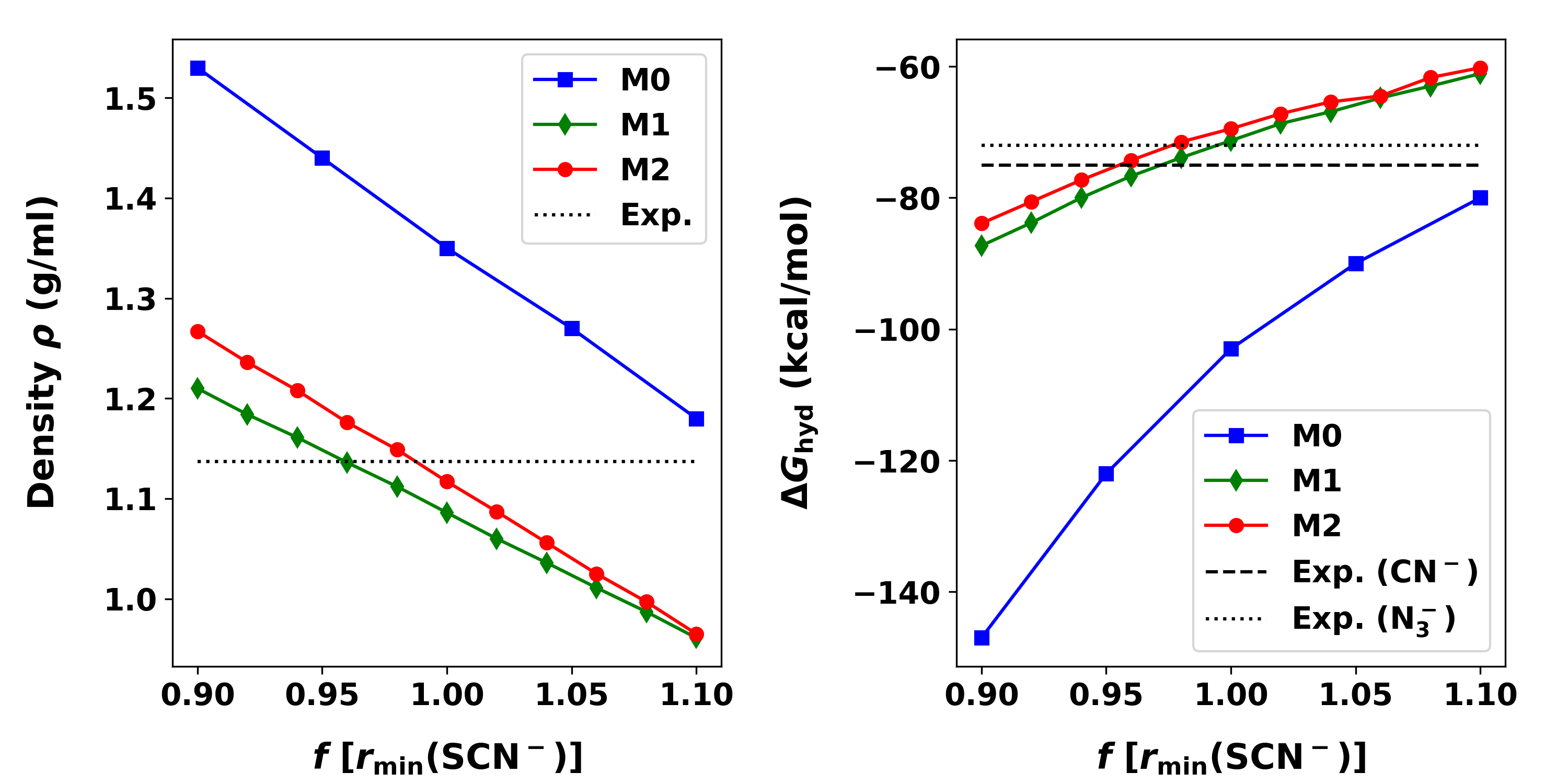}
\caption{Left panel: Computed density of a $3.8$\,mol/kg KSCN solution
  in water at different scaled $r_\mathrm{min}$ values using models
  {\bf M0} (scaled LJ parameters\cite{bian:2013} with MTP
  electrostatics, blue), {\bf M1} (scaled LJ
  parameters\cite{lund:2018} with fMDCM, green) and {\bf M2} (cluster
  fitted LJ parameters with fMDCM, red). Right panel: Hydration free
  energies $\Delta G_\mathrm{hyd}$ of a single SCN$^-$ ion in water
  solution. The dashed lines are experimental hydration free energies
  for N$_3^{-}$ and CN$^{-}$ and the reported $\Delta G_{\rm hyd} \sim
  -72$ kcal/mol for SCN$^{-}$.\cite{marcus:1997}}
\label{fig_rho_hfe}
\end{figure}

\noindent
For model {\bf M0} using the MTP electrostatic model, the
literature\cite{bian:2013} LJ parameters $r_\mathrm{min}$ of the
SCN$^-$ atoms were scaled by multiplying with a factor $f_\mathbf{M0}=
\{0.9, 0.95, 1.0, 1.05, 1.1 \}$. Using these scaled parameters, the
system density and hydration free energy of SCN$^-$ were determined,
respectively. Figure \ref{fig_rho_hfe} (blue lines) reports the
variation of $\rho$ and $\Delta G_{\rm hyd}$ with scaling $f$ and
provides an estimate for $f_{\rm M0}$ to best reproduce the two
experiments. With $f_\mathbf{M0} = 1.1$ the average density is
$\rho_\mathrm{M0} = 1.18$\,g/cm$^3$ and the hydration free energy of
SCN$^-$ in water is $\Delta G_{\mathrm{hyd},\mathbf{M0}} =
-80$\,kcal/mol. These are significant improvements of 10\% and 20\%,
respectively, over the two reference values for $f = 1.0$ (i.e. using
the original LJ parameters).\\

\noindent
Model {\bf M1} (fMDCM electrostatics) was based on LJ parameters that
had been used for modeling aqueous SCN$^-$ solutions.\cite{lund:2018}
To adjust the literature LJ parameters to the fMDCM electrostatic
representation, $r_\mathrm{min}$ values - which are equivalent to
$\sigma$ by way of $r_\mathrm{min} = \sqrt[6]{2}\sigma$ - of the
SCN$^-$ atoms were again scaled by a factor $f_\mathbf{M1}$ to best
reproduce the experimentally measured observables. Figure
\ref{fig_rho_hfe} (green lines) reports $\rho$ and $\Delta
G_\mathrm{hyd}$ for different scaling factors in a scan range of
$f=[0.9,1.1]$ and the experimental reference (horizontal black dotted
and dashed lines). For model {\bf M1} and $f_\mathbf{M1} = 0.96$, the
computed density $\rho_\mathbf{M1} = 1.135$\,g/cm$^3$ compares well
with the experimentally measured\cite{albright:1992} density
$\rho_\mathrm{exp}=1.139$\,g/cm$^3$ and the computed hydration free
energies $\Delta G_{\mathrm{hyd},\mathbf{M1}} = -76.7$\,kcal/mol
($-71.3$\,kcal/mol for unscaled LJ parameter) is within reasonable
range of the experiments. \\

\begin{figure}
  \centering
  \includegraphics[width=0.75\textwidth]{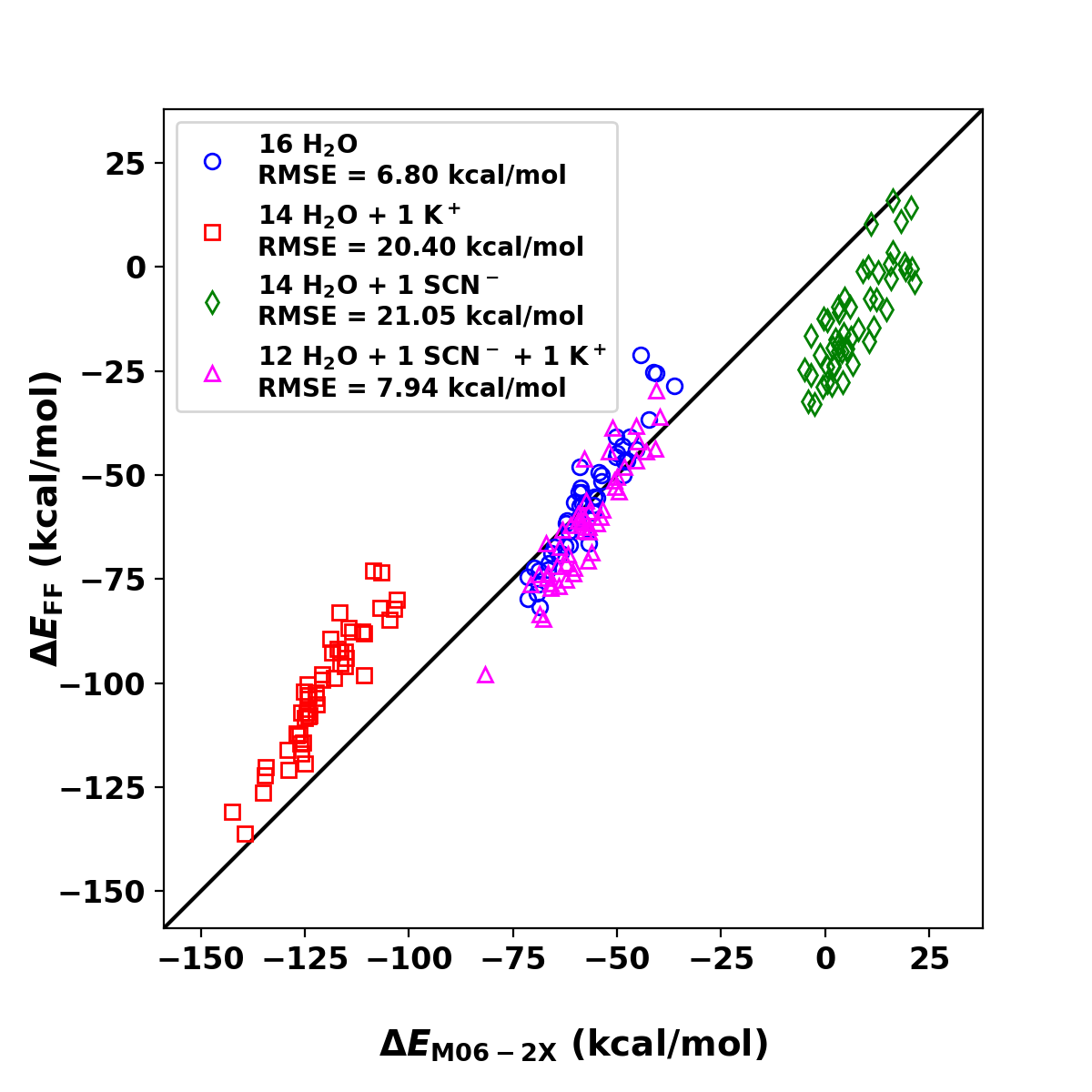}
\caption{Correlation between interaction energies from reference {\it
    ab initio} calculations and those based on the fitted model {\bf
    M2}. Random snapshots were taken from equilibrium simulations. The
  systems considered always consist of one SCN$^-$ anion surrounded by
  shells as indicated in the legend with corresponding RMSE. The
  overall RMSE is $15.56$\,kcal/mol.}
\label{fig_ecorr}
\end{figure}

\noindent
The interaction energy consists of electrostatic contributions using
the fMDCM model for SCN$^-$ and the vdw interactions described as a LJ
potential. The optimized LJ parameters achieved an overall RMSE of
$15.6$\,kcal/mol for all snapshots. The individual RMSE values are (i)
$8.3$\,kcal/mol, (ii) $20.4$\,kcal/mol, (iii) $20.5$\,kcal/mol, and
(iv) $9.5$\,kcal/mol. The reference interaction energies cover a broad
range from $\sim -142$\,kcal/mol to $\sim +18$\,kcal/mol, see Figure
\ref{fig_ecorr}. In a second step, the optimized $r_\mathrm{min}$ of
the SCN$^-$ atoms were again scaled by a factor $f$ and the density of
an aqueous KSCN solution and hydration free energy $\Delta
G_{\mathrm{hyd}}$ were computed from explicit MD simulations, see
Figure \ref{fig_rho_hfe} (red lines). Closest agreement with
  experiment was found for $f_\mathbf{M2} = 0.98$, very close to $f=1$
  (no modifications in the van der Waals parameters), which yields
$\rho_\mathbf{M2} = 1.119$\,g/cm$^3$ and $\Delta
G_{\mathrm{hyd},\mathbf{M2}} = -73.0$\,kcal/mol. Hence, fitting the
interaction energies for a diverse set of hydrated structures yields
LJ parameters that are suitable for thermodynamic observables from
simulations.\\

\noindent
In conclusion, all three models allow to adjust the Lennard-Jones
parameters for MTP and fMDCM electrostatics such as to satisfactorily
reproduce experimentally measured thermodynamic reference data, see
Figure \ref{fig_rho_hfe}. The actual scaling parameters are close to
$f=1$ for the fMDCM models ($f_\mathbf{M1} = 0.96$, $f_\mathbf{M2} =
0.98$) and larger by $\sim 10$\% for the van der Waals ranges if MTP
($f_\mathbf{M0} = 1.1$) is used. Scaling factors of $f \sim 1.1$ for
MTP-based models are consistent with earlier work on the vibrational
relaxation and hydration free energies of CN$^{-}$ and
OH$^{-}$.\cite{MM.cn:2013}\\

\subsection{Structure and Ordering of the Mixture}
\begin{figure}
  \centering
  \includegraphics[width=0.95\textwidth]{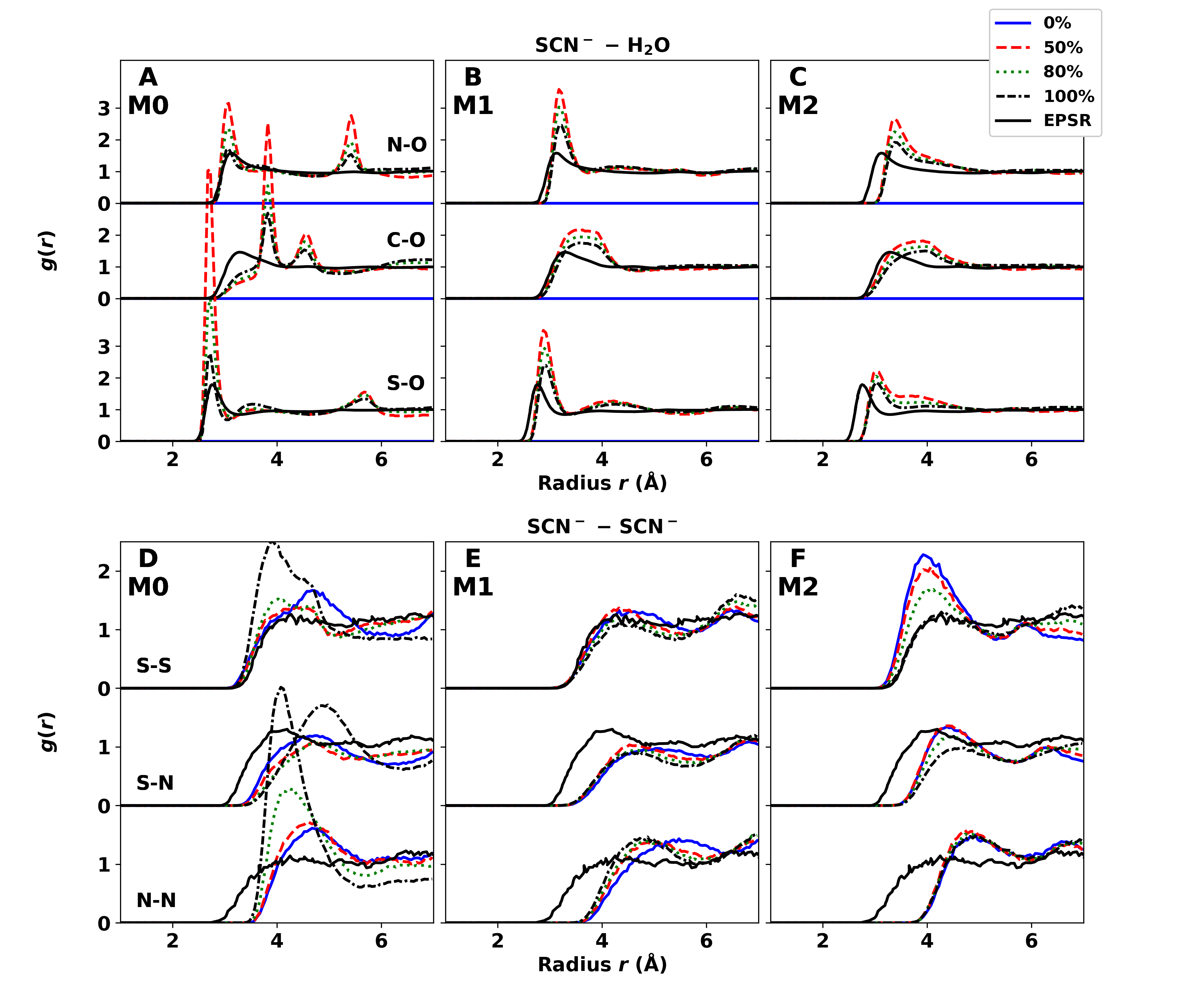}
\caption{Radial distribution function $g(r)$ between (A-C) SCN$^-$
  atoms and oxygen of water and (D-F) between both SCN$^-$ anions.
  The results are shown from simulations using (A, D) the MTP model
  and scaled LJ parameters {\bf M0}, (B, E) the fMDCM approach and
  scaled LJ parameters {\bf M1} and (C, F) the fMDCM approach and
  fitted LJ parameters {\bf M2}. In comparison to the $3.8$\,mol/kg
  KSCN/water mixture (100\%), $g(r)$ of an aqueous KSCN solution
  ($2.5$\,mol/kg) obtained from empirical potential structure
  refinement (EPSR) to match experimentally measured neutron
  diffraction data are shown as solid black lines.\cite{botti:2009}}
\label{fig_g_fdcm}
\end{figure}

Using the optimized LJ parameters for models {\bf M0} to {\bf M2},
equilibrium MD simulations 25\,ns in length were carried out for each
system composition. From these simulations, radial pair distribution
functions $g(r)$ were determined and compared with results from
empirical potential structure refinement (EPSR) fits to best reproduce
neutron diffraction measurements of aqueous KSCN
solution.\cite{botti:2009} It is important to stress that such pair
distribution functions are not measured directly from experiments but
rather adjusted empirically to reproduce the neutron scattering
amplitude which is written as a weighted average over pair
distribution functions.\cite{botti:2009} Figure \ref{fig_g_fdcm}
reports SCN$^{-}$--water (top) and SCN$^{-}$--SCN$^{-}$ (bottom) pair
correlation functions for different mixtures KSCN in solution (0,
50\%, 80\%, 100\% water content) compared with experimentally obtained
$g(r)$. The radial distribution functions $g(r)$ in Figures
\ref{fig_g_fdcm}B and C (models {\bf M1}, {\bf M2}) are considerably
smoother than those using the MTP model in Figure \ref{fig_g_fdcm}A
({\bf M0}). For simulations in pure water (100\%) direct comparisons
between simulations (dot-dashed black line) and EPSR results (solid
black line) are possible.\cite{botti:2009} For model {\bf M0}, the
overall shape of $g_{\rm N-O_{W}}(r)$ agrees reasonably well with the
EPSR results although the computed pair distribution function is
overstructured with a local maximum around 5.5\,\AA. Such maxima are
not found for models {\bf M1} and {\bf M2} for which the position of
the maximum is shifted to larger separations and the height of the
maximum is overestimated for {\bf M1} but quite well described by {\bf
  M2}. With decreasing water content the height of the first peak
increases. In other words, with decreasing water density the ion
  recruits water molecules. The shift of the first maximum for model
{\bf M2} relative to model {\bf M1} is consistent with the larger
$r_{\rm min}$ values for the S- and N-atoms for {\bf M2} compared with
{\bf M1}, see Tables S3 and
S4.\\

\noindent
For the C--O$_{\rm W}$ and S--O$_{\rm W}$ pair correlation functions
$g_{\rm C-O_W}(r)$ and $g_{\rm S-O_W}(r)$, similar observations are
made. In all cases, the pair distribution functions from model {\bf
M0} are overstructured with additional peaks at larger separations
(particularly for $g_{\rm C-O_W}(r)$). On the other hand, radial
distribution functions from simulations using models {\bf M1} and 
{\bf M2} are in reasonably good agreement with EPSR results - in
particular for large separations - whereas for the first solvation
shell the position of the first maximum is at larger separations and
the peak heights for {\bf M2} are closer to the EPSR results.\\

\noindent
The difference in the peak intensities for $g(r)$ obtained from the
MTP and fMDCM simulation indicates favouring considerably stronger
binding of H$_2$O to SCN$^-$ ions by the MTP model. This also leads to
a more rigid second solvation shell indicated by significant peak
heights $g(r) \gtrsim 5$\,\AA at larger values of $r$ and is
consistent with the increased hydration free energies even after
scaling the $r_{\rm min}$ values by $f_{\rm M0} = 1.1$, see Figure
\ref{fig_rho_hfe}.  \\

For simulations of KSCN in water/acetamide solvent mixtures with fMDCM
(models {\bf M1}, {\bf M2}), the peak intensities in $g(r)$ between
SCN$^-$ atoms and water oxygen increases for larger acetamide
concentration (with exception of the water-free system 0\%) but the
peak positions remain unchanged. A similar observation for different
mixtures is made for simulation with model ({\bf M0}), but with
considerably more pronounced increases in the peak intensities with
larger acetamide content in the solvent mixture.  \\

\noindent
For anion-anion pair distribution functions (Figures \ref{fig_g_fdcm}D
to F) the agreement for $g_{\rm S-S}(r)$ for models {\bf M1} and {\bf
  M2} is particularly notable. Notwithstanding, model {\bf M0} is also
capable of qualitatively describing the experiments. For the S--N and
N--N pair correlation functions model {\bf M0} has the first maximum
either shifted to longer separations $r$ or a much too pronounced
maximum.  Compared with that, models {\bf M1} and {\bf M2}
  correctly capture the peak height for $g_{\rm S-N}(r)$ and $g_{\rm
    N-N}(r)$, which is a proxy for the interaction strength.  The
  position of the first maximum, however, is at larger separation as
  seen for model {\bf M0}.  Opposite to the reference EPSR results,
  the position of the first maximum in $g_{\rm N-N}(r)$ is at larger
  separation than in $g_{\rm S-S}(r)$ due to a higher concentration of
  negative charge on the N- rather than the S-atom in SCN$^-$
  according to the fitted fMDCM ($\sum q_{i, \rm N} = -1.142$\,e
  vs. $\sum q_{i, \rm S} = -0.858$\,e) and MTPL ($q_{\rm N} =
  -0.455$\,e vs. $q_{\rm s} = -0.182$\,e) models.  \\

\noindent
It should
be noted that the peak position of the anion-anion pair distribution
functions in simulations can be steered by adjusting the LJ parameter
$r_\mathrm{min}$ of the SCN$^-$ atoms individually. However, the
procedure followed here was not to match pair distribution functions
$g(r)$ as the LJ parameters of SCN$^-$ are at first either taken from
literature ({\bf M0}, {\bf M1}) or fit to {\it ab initio} data ({\bf
  M2}) and later scaled collectively to best match the density of an
aqueous KSCN system or hydration free energy of SCN$^-$ in water.  \\

\noindent
Peak intensities for SCN$^-$ atom pairs in different water/acetamide
solvent mixture remains either similar ({\bf M1}) or increase for
larger acetamide content ({\bf M2}). For simulations with multipoles
(model {\bf M0}), the peak positions are generally closer those
reported from experiments but with higher intensities which are
highest for pure water solvent and decreases for larger acetamide
content in the water/acetamide solvent mixture. \\

\begin{figure}
  \centering
  \includegraphics[width=0.80\textwidth]{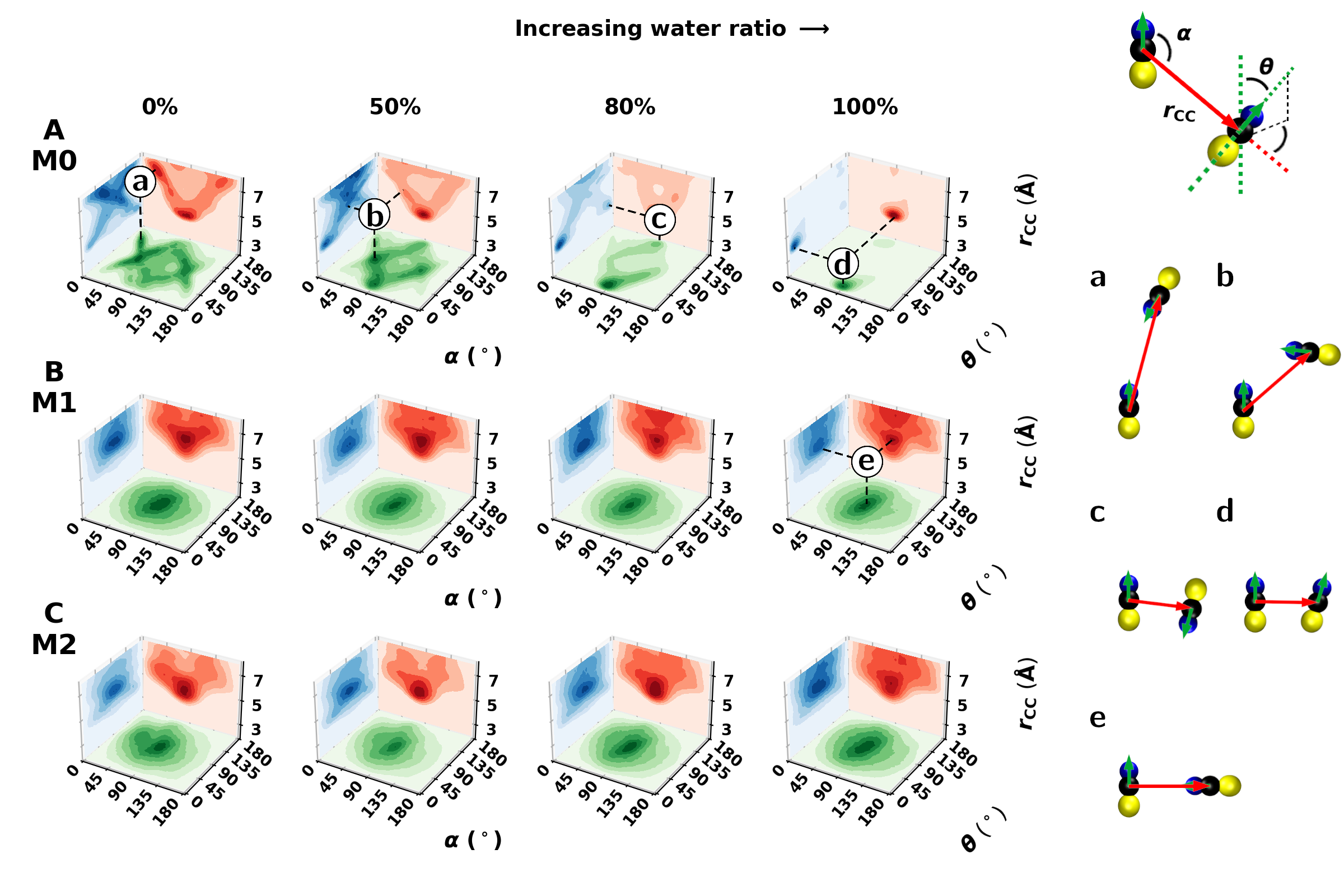}
\caption{Radial-angular distributions of SCN$^-$ pairs in different
  water/acetamide solutions (columns) from simulations with different
  models for the intermolecular interactions (rows). Specific
    features of the distributions are labelled as encircled a to e. The top right
    figure defines the coordinates used and sketches a to e below illustrate
    typical geometries corresponding to the features a to e in the main figure.}
\label{fig_radang}
\end{figure}

\begin{figure}
  \centering
  \includegraphics[width=0.99\textwidth]{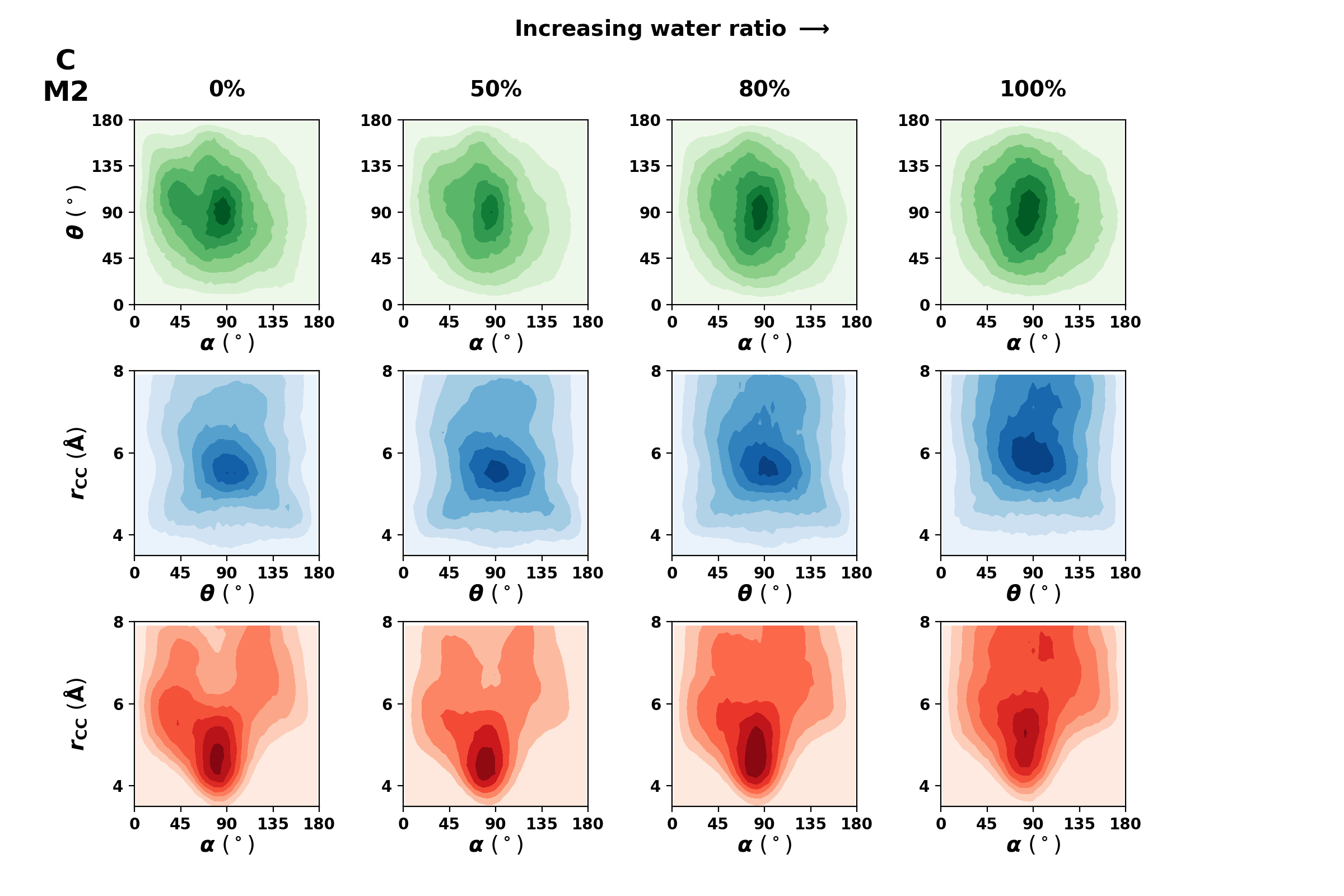}
\caption{Contour plots of the radial-angular distribution plots of
  SCN$^-$ pairs in different water/acetamide solutions (columns) from
  simulations with the setup {\bf M2}.}
\label{fig_radang2}
\end{figure}

\noindent
Figure \ref{fig_radang} reports radial-angular pair distribution
  functions for SCN$^-$/SCN$^-$ pairs along the C-C bond distances
and (CN)-(CN) bond angles (related to SCN$^-$ dipole vectors). Even
after adjusting the LJ parameters for {\bf M0} to yield improved
thermodynamics, the MTP-based model lead to mainly angled SCN$^-$
pairs in pure acetamide and solvent mixtures with large acetamide
content. This changes to predominantly parallel and, partly,
anti-parallel SCN$^-$ pair alignments with increasing water content
all of which is consistent with results from simulations using the
unmodified MTP-based model.\cite{MM.eutectic:2022} In simulations
using fMDCM for the electrostatics (models {\bf M1} and {\bf M2}), the
angular pair SCN$^-$ distribution functions depend less strongly on
solvent mixtures and are close to a sine-angle distribution
  which is the probability to find a particle at a certain
  angle on a sphere ($P(\theta, \phi) = | Y_{0,0} | \sin \theta$).
  This can be even better seen in
Figure \ref{fig_radang2} which reports the 2-dimensional projections
of the distributions shown in Figure \ref{fig_radang} using model {\bf
  M2}.\\

\noindent
Contrary to {\bf M0}, the probability distributions do not change so
profoundly with increasing water content for models {\bf M1} and {\bf
  M2}. Nevertheless, the centroid of the C--C probability
distributions shift to somewhat longer C--C separations with
increasing water content. Also, the maximum for $P(r_{\rm CC},\alpha)$
is more sharply peaked for low water content and broadens around the
maximum for pure water as the environment. Conversely, the two
higher-probability regions at large C--C separations $\sim 7$ \AA\/
that are present with acetamide as the solvent disappear with
  increasing water content. \\

\subsection{Dynamics of the Mixtures}
\begin{figure}
\centering
\includegraphics[width=0.90\textwidth]{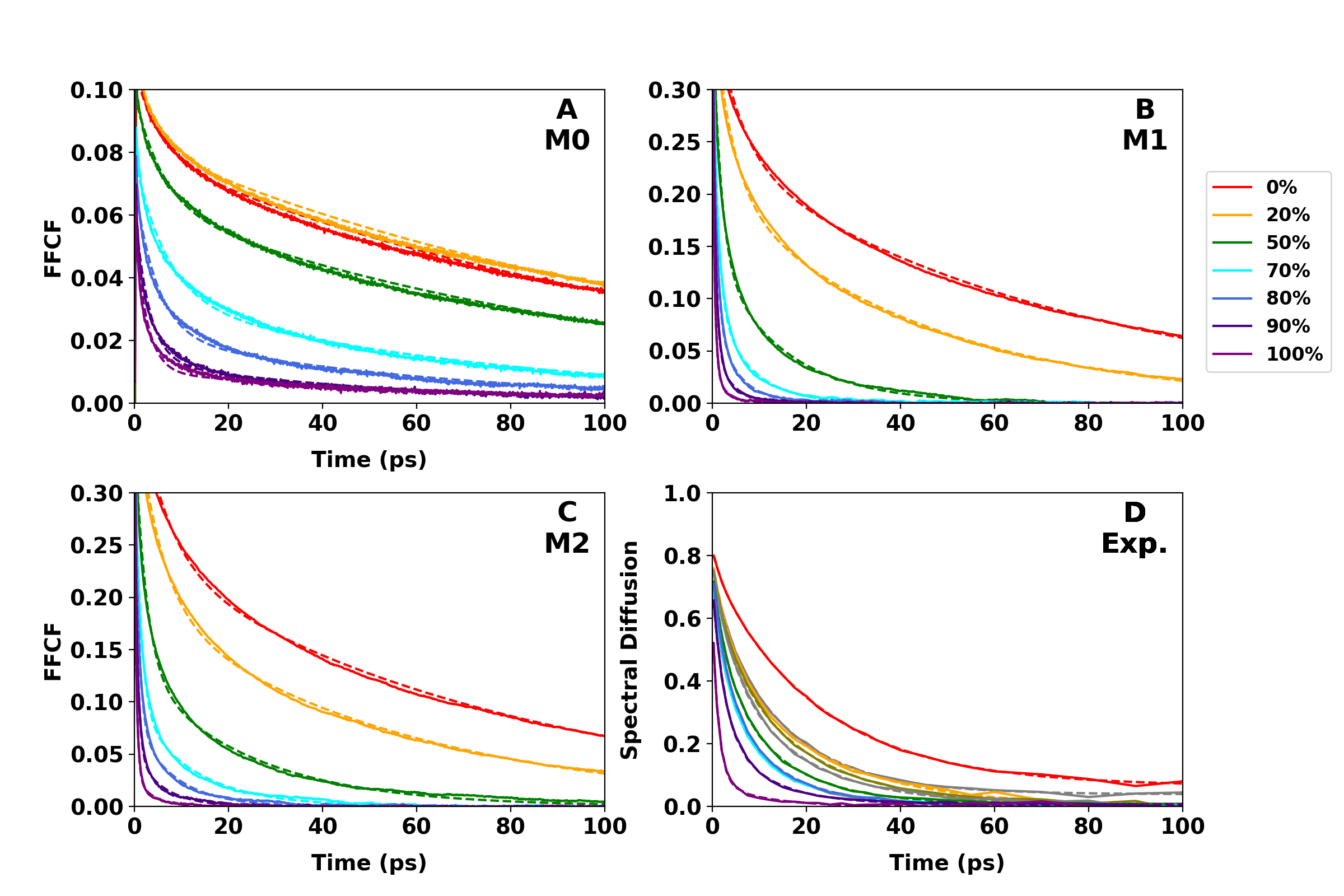}
\caption{Panels A to C: FFCF (solid lines) of the INM frequencies
  $\nu_3$ of SCN$^-$ anion from simulation with different force fields
  for different mixing ratios of acetamide and water. Panel D:
  Experimentally measured spectral diffusion.  Gray lines in
    panel D are experimental results for which no simulations were
    carried out. A bi-exponential function (dashed lines) was fit to
  each computed FFCF (A-C) and to the experimental data in panel D in
  the range from $0.25$\,ps to 1\,ns. All computed FFCFs are
  normalized.}
\label{fig_ffcf}
\end{figure}

Finally, the dynamics of the mixtures was studied by analyzing the
C--N vibrations from instantaneous normal modes of the SCN$^-$ anion
and the corresponding frequency fluctuation correlation functions
(FFCF) for varying water content. Figure \ref{fig_ffcf}A to C shows
the FFCFs with models {\bf M0} to {\bf M2}, respectively. Figure
\ref{fig_ffcf}D reports the experimentally measured FFCFs from the
2D-IR experiments. It is found that the water content affects the
FFCFs in a pronounced fashion: with increasing water content the fast
and slow time scales decrease. For the FFCFs from the simulations a
bi-exponential fit is a meaningful model applicable to all
mixtures. Assuming a single exponential significantly
  deteriorates the fit. To allow direct comparison between experiment
  and simulations, the experimental data was also fit to a
  double-exponential. The set of optimized parameters of the
bi-exponential fit function are shown in Figure S3
and Table S6.  Qualitatively, models {\bf M1}
  and {\bf M2} capture the experimentally observed FFCFs and their
  dependence on the amount of water in the solvent mixture whereas
  model {\bf M0} does not. Overall, the decay on the longer
  time scale from simulations using {\bf M0} is too slow. However,
the relative amount by which the time constant changes between 0\% and
100 \% water is considerably closer to the experimentally observed
change than in the other two models.\\

\begin{figure}
\centering \includegraphics[width=0.90\textwidth]{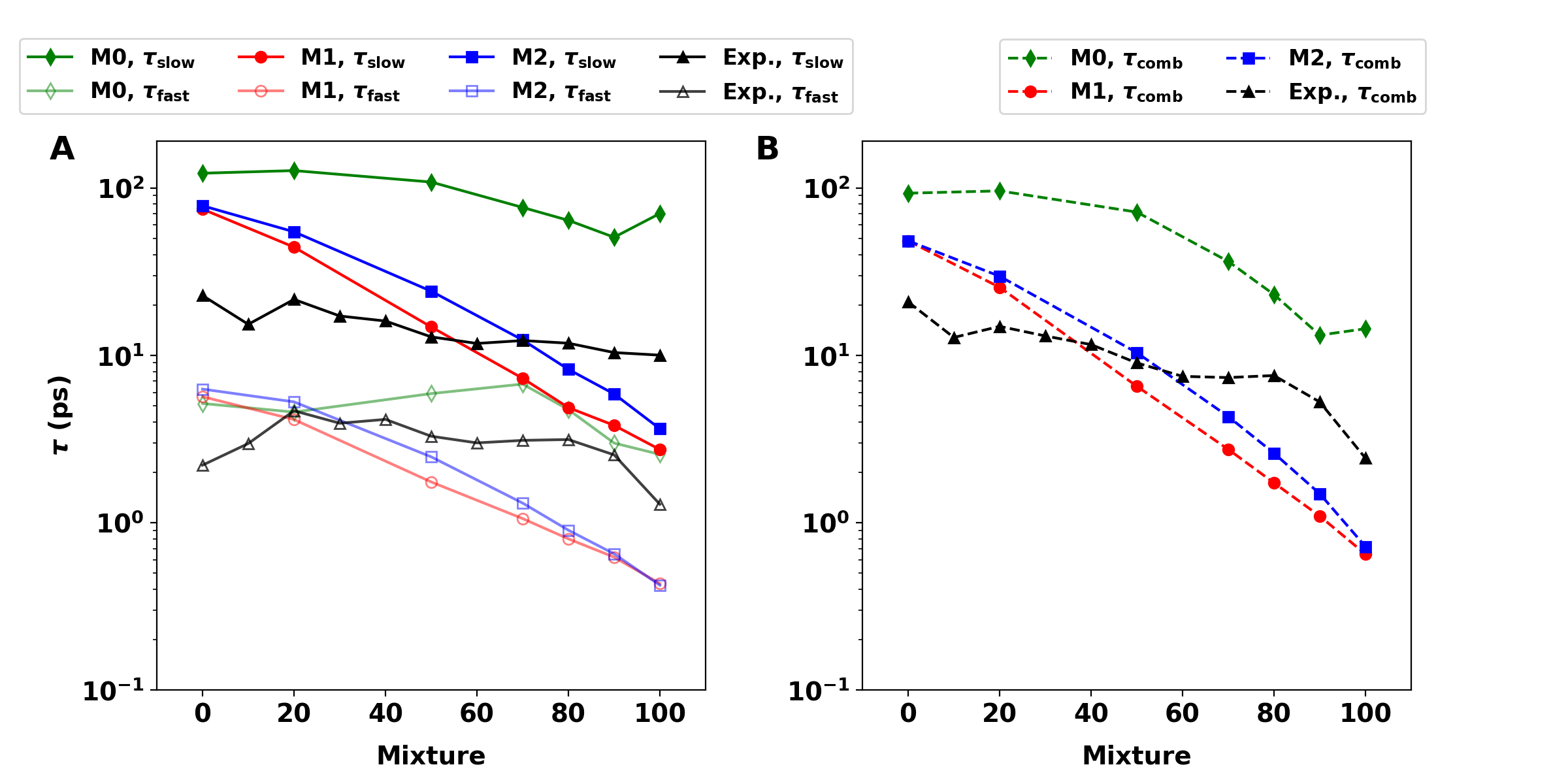}
\caption{Panel A: Fitted lifetimes $\tau_\mathrm{slow}$ (full marker)
  and $\tau_\mathrm{fast}$ (open marker) of a bi-exponential function
  to the FFCF of the INM frequencies $\nu_3$ of the SCN$^-$ anion from
  simulation with models {\bf M0}, {\bf M1} and {\bf M2}. The
  lifetimes from the bi-exponential fit to the experimental spectral
  diffusion are shown in black. Panel B: Amplitude-weighted decay
  times $\tau_{\rm comb}$ for all three models, see text.}
\label{fig_times}
\end{figure}

\noindent
A more quantitative analysis is based on the decay times of the FFCFs
depending on water content, see Figure
\ref{fig_times}. Experimentally, the slow decay $\tau_{\rm slow}^{\rm
  exp}$ (black filled triangles) almost monotonously decreases with
increasing water content. This is qualitatively reproduced from
simulations using all three models although the amplitude from model
{\bf M0} (green filled diamonds) is larger by a factor of $\sim 6$ and
those from models {\bf M1} and {\bf M2} (red filled circles and blue
filled squares) are larger by a factor of $\sim 4$ and the slope is
considerably steeper. For the fast decay $\tau_{\rm fast}^{\rm exp}$
(black open triangles) an increase with increasing water content up to
$\sim 40$ \% is observed followed by a monotonous decrease. This is
qualitatively captured by model {\bf M0} (green open diamonds),
although the maximum is shifted to higher water content, whereas
models {\bf M1} and {\bf M2} (red open circles and blue open squares)
again find monotonous decrease to amplitudes considerably lower than
the experiment for pure water.\\

\noindent
However, it is known that in biexponential fits the final parameters
can be strongly correlated. Therefore, it may be more appropriate to
compare, for example, amplitude-weighted decay times $\tau_{\rm comb}
= a_{\rm slow} \tau_{\rm slow} + a_{\rm fast} \tau_{\rm fast}$, see
Figure \ref{fig_times}B, with amplitudes reported in Figure
S3A. Overall, the computations reproduce the
decrease in $\tau_{\rm comb}$ with increasing water content. For {\bf
  M0} the experimentally determined values are typically overestimated
by an order of magnitude or less. This differs for models {\bf M1} and
{\bf M2} which overestimate $\tau_{\rm comb}$ for low water content
and underestimate it for high water content. Specifically, for the
water-free system (0 \%) $\tau_{\rm comb}^{\rm exp} \sim 21$ ps which
compares with $\langle \tau_{\rm comb} \rangle \sim 48$\,ps from
models {\bf M1} and {\bf M2} and $\langle \tau_{\rm comb} \rangle \sim
100$\,ps. Together with the results from Figure \ref{fig_ffcf} it is
concluded that the two fMDCM-based models more realistically describe
the solvent dynamics as encapsulated in the FFCFs.\\

\section{Discussion and Conclusion}
The present work improves, validates and applies an energy function
for a heterogeneous eutectic mixture consisting of water, ions and
acetamide. For the SCN$^-$ ion a flexible MDCM model was
developed. The van-der-Waals parameters were optimized vis-a-vis
experimental densities and hydration free energies using the TIP3P
water model for consistency with the remaining force field
parameters. This point will be considered further below. The bonded
terms for SCN$^-$ were replaced by an accurate reproducing kernel
representation of reference calculations at the
PNO-LCCSD(T)-F12/aug-cc-pVTZ-F12 level of theory. Three different
parametrization approaches were pursued and all of them describe the
experimental reference data satisfactorily.\\

\noindent
Force fields for deep eutectic solvents have been previously optimized
using different protocols and
approaches\cite{ferreira:2016,padua:2022,
  maglia:2021,doherty:2018,jeong:2021,garcia:2015,zhang:2022,velez:2022}
Most common to these efforts is to start from a conventional energy
function such as the General Amber Force Field (GAFF),
\cite{wang:2004} adapted for particular
applications\cite{perkins:2014} and to further readjust
parameters.\cite{ferreira:2016,zhang:2022} The adjustment is usually
to scale partial charges\cite{garcia:2015} and in some cases
van-der-Waals parameters to reproduce observed properties such as
diffusivities, viscosities or the densities.\cite{zhang:2022} In terms
of using diffusivities for the parametrization from explicit MD
simulations it is noted that in DES they are several magnitudes slower
than those in typical liquids such as water ($10^{-11}$ to $10^{-12}$
m$^2$/2 vs. $10^{-9}$
m$^2$/s).\cite{jeong:2021,ferreira:2016,nilsson:2001} Thus the
determination of self-diffusion coefficients $D$ typically requires a set of $\sim
50$\,ns simulations (compared with 5 ns for water\cite{nilsson:2001})
which would make an explicit parametrization extremely expensive.
Alternatively, there have been efforts to develop models based on
symmetry adapted perturbation theory (SAPT)\cite{jeong:2021} which
were refined by comparing with first principles MD
simulations. Another approach was chosen for the Canongia Lopes \&
Padua (CL\&P) force field and its polarizable variant
CL\&Pol.\cite{padua:2022} Recent fine-tuning efforts used structural
data.\cite{maglia:2021} The present status of computational approaches
for deep eutectic solvents has been recently
reviewed.\cite{velez:2022}\\

\noindent
In the present work two alternative procedures were followed, both
based on improved electrostatics for the SCN$^{-}$ anion and
subsequently adjusting the LJ parameters. First, a previously used
distributed multipole model was improved by scaling the LJ parameters
to correctly describe experimentally measured liquid densities and
hydration free energies (model {\bf M0}). To probe the influence of
the electrostatic model used, a fluctuating MDCM model was constructed
to replace the multipoles, and the LJ parameters were adjusted as for
the multipolar model (model {\bf M1}). Second, reference interaction
energies from hydrated SCN$^{-}$ clusters of different compositions
were determined from electronic structure calculations (model {\bf
  M2}). After subtracting the electrostatic contributions (fMDCM for
SCN$^{-}$) the LJ parameters were scaled to best describe the
remaining nonbonded interactions. Finally, a second round of
refinement by scaling the LJ $r_\mathrm{min}$ parameters of SCN$^{-}$
followed to best describe again experimental liquid densities and
hydration free energies as for {\bf M0} and {\bf M1}. This
  strategy follows common practices in empirical energy function
  development in that the electrostatic parameters usually originate
  from high-level quantum chemical calculations whereas the
  van-der-Waals parameters are adjusted with respect to, for example, interaction
  energies and hydration free energies.\cite{mackerell:2024}\\

\noindent
Interestingly, the optimal scaling parameter for model {\bf M2} to
best describe experimental observables was $f_\mathbf{\rm M2} =
1.0$. This suggests that fitting LJ parameters to interaction energies
from sufficiently high-level electronic structure calculations and
using a robust charge representation for the ion (fMDCM) yields models
that can be used for meaningful simulation of thermodynamic
observables. It will be interesting to see how generic such an
approach is by applying it to other, similar systems. On the other
hand, model {\bf M0} required rather large changes in the LJ
parameters (up to 10 \%) to obtain agreement with the thermodynamic
data considered here. This is consistent with earlier work which also
found that in particular LJ $r_\mathrm{min}$ parameters (or
$r_\mathrm{min} = \sqrt[6]{2}\sigma$)) needed to be increased by $\sim
10$ \% for hydration free energies consistent with
experiment.\cite{MM.cn:2013}\\

\noindent
Decomposition of a molecular ESP into atomic distributed multipoles
can be viewed as a linear combination of atom-centered spherical
harmonics. The expansion coefficients are the atomic charge, dipole,
and higher order multipole moments. Usually, this series is truncated
at the atomic quadrupole. The basis functions of this expansion
(``atomic orbitals'') have a rather pronounced directionality which
influences short- to mid-range intermolecular interactions. Contrary
to that, point charge-based models such as fMDCM represent the
molecular ESP as a superposition of isotropic potentials. This can be
expected to lead to less pronounced angular variations. Both models -
multipole- and PC-based - can be fit to the same quality compared with
reference molecular ESPs from electronic structure
calculations. Qualitatively, the increased LJ $r_\mathrm{min}$
parameters required for model {\bf M0} (MTP-based) compared with {\bf
  M1} (PC-based) indicate that the more pronounced anisotropies from
atom-based multipoles need to be damped to moderate interactions with
nonbonded partners and to potentially reduce the pronounced
anisotropy. \\

\noindent
The changes in the LJ parameters for model {\bf M0} also influence
measured spectroscopic properties. For example, the vibrational energy
transfer between neighboring SCN$^{-}$, which was measured
previously,\cite{MM.eutectic:2022} is affected by intermolecular
coupling and structural alignment. The more pronounced anisotropy of a
MTP-based representation is also seen in the 3-dimensional pair
distribution functions $P(\alpha,\theta,r_{\rm CC})$, see Figure
\ref{fig_radang}.  For model {\bf M0} the angular dependence of these
distribution functions is considerably more structured compared with
models {\bf M1} and {\bf M2}. It is also interesting to note that
despite increasing the LJ $r_\mathrm{min}$ by $\sim 10$ \% the
$P(\alpha,\theta,r_{\rm CC})$ for {\bf M0} from the present work
compared with the earlier employed
parametrization\cite{MM.eutectic:2022} do not differ appreciably
(compare with Figure 5 in Ref.\cite{MM.eutectic:2022})\\

\noindent
Using the three interaction models the viscosities for varying
acetamide/water ratios were determined as an additional validation,
see Figure S4. For this, the stress tensor
$\mathbf{P}(t)$ was computed and the viscosity was determined from
$\eta = \frac{V}{6 k_B T} \sum_{\alpha \le \beta} \int_{0}^{\infty}
\langle \bar{P}_{\alpha\beta}(0) \bar{P}_{\alpha\beta}(t) \rangle
\mathrm{d}t$ $(\alpha, \beta = x, y, z)$ where $\bar{P}_{\alpha\beta}$
are the upper triangular elements of the modified stress tensor
$\mathbf{\bar{P}}(t)$.\cite{visc_zhang:2020,zhang:2022} Due to the
strong intermolecular interactions and high viscosities, converging
$\eta$ can be rather demanding and is not attempted here. Rather, 5
independent $NVT$ simulations were carried out for each composition
and the results were averaged to obtain illustrative results.\\

\noindent
For model {\bf M0} the viscosity first fluctuates around $\sim 300$
mPa$\cdot$s before decaying with increasing water content (Figure
S4A). On the other hand, models {\bf M1} and {\bf
  M2} find that $\eta$ monotonously decreases with increasing water
content, see Figures S4B and C. The maximum
viscosities, which occur for zero water content, differ by a factor of
two. Experiments for the water-free system report $\eta \sim 100$
mPa$\cdot$s at 303\,K.\cite{liu:2013} Simulations with {\bf M1} and
{\bf M2} are consistent with such values. More importantly, however,
the experiments report that $\eta$ decreases monotonously with
increasing water content what is found by models {\bf M1} and {\bf M2}
but not {\bf M0}. The magnitude of $\eta$ for the three models also
reflects the strength of the intermolecular interactions. Given the
large scatter in $\eta$ from individual simulations, in particular for
low water-content, reaching convergence requires longer or more
simulation data. From a computer efficiency perspective it appears to
be advisable to consider $\eta$ as a validation property rather than a
property used for fitting the intermolecular interactions.\\

\begin{figure}
  \centering
  \includegraphics[width=0.50\textwidth]{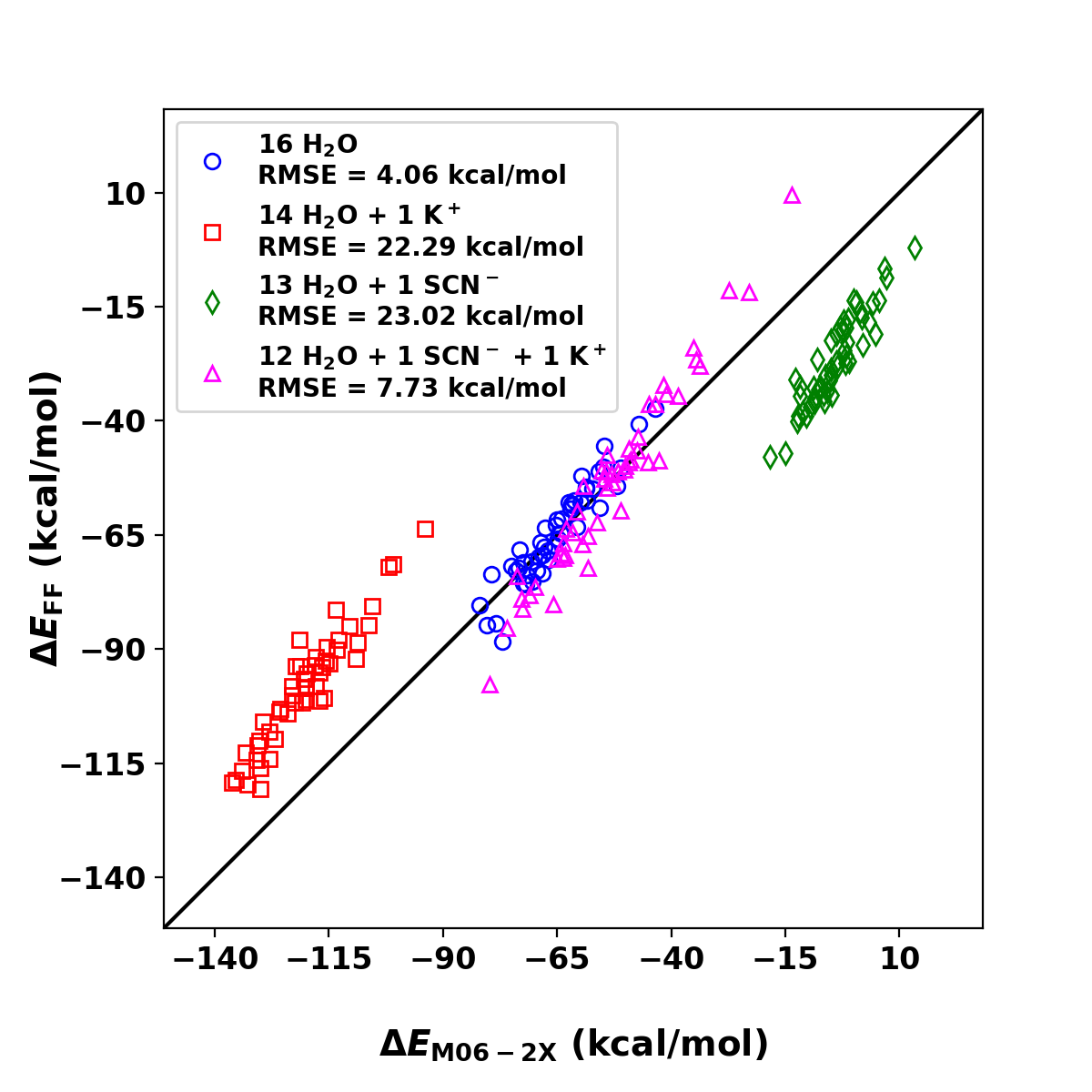}
\caption{Correlation between interaction energies of SCN$^-$ anion a
  surrounded by shells of different compilation computed by reference
  {\it ab initio} calculations and a fitted model setup equivalent to
  {\bf M2} but with TIP4P water residues instead of TIP3P. Random
  snapshots were taken from equilibrium simulations of KSCN in
  TIP4P. The systems considered always consist of one SCN$^-$ anion
  surrounded by shells as indicated in the legend with corresponding
  RMSE. The overall RMSE is $16.61$\,kcal/mol.}
\label{fig_tip4p}
\end{figure}

\noindent
As a final comment, the TIP3P water model used here is known to have
various limitations when comparing experimentally measured quantities
with those from simulations.\cite{kadaoluwa:2021} As was already
mentioned, the reason to use the TIP3P model was consistency with the
CGenFF parameters for the other species present in the eutectic
mixture. To quantitatively assess changes using a different water
model, MD simulations were carried out using model {\bf M2} but with
the TIP4P water model.\cite{TIP3P-Jorgensen-1983} Again, hydrated
SCN$^{-}$ clusters were extracted, reference interaction energies
computed, and the parametrization step was repeated as had been done
for TIP3P, see Figure \ref{fig_ecorr}. The results are reported in
Figure \ref{fig_tip4p} and Table S4.  It is
found that the two water models yield similarly accurate
representations of the intermolecular interactions. With this new
parametrization, simulations using the TIP4P model can be carried out
to determine possible changes for the structure and spectroscopy when
using a water model different from TIP3P. This, however, is beyond the
scope of the present work. Another future extension of the present work is to use kernel-based MDCM (kMDCM) to capture conformationally dependent changes in the ESP in particular for molecules larger than triatomics.\cite{MM.kmdcm:2024}\\

\noindent
In summary, the parametrization strategies pursued in the present
  work successfully combined electronic structure methods, comparison
  with thermodynamic reference data and application to measured
  structural and spectroscopic properties. Adjustment of both,
  electrostatic contributions and LJ-parameters was found to be
  important. The present work suggests that distributed charge models
  are preferable over multipole-based representations of the
  electrostatics for the system compositions considered
  here. Application of the procedures outlined here to other, related
  mixtures and additional observables will provide further insight
  into generalizability and scope for improvements of this
  approach.\\

\section*{Acknowledgment}
This work has been financially supported by the Swiss National Science
Foundation (NCCR MUST, 200020\_219779, 200021\_215088 to MM), the
University of Basel (to MM) by European Union's Horizon 2020 research
and innovation program under the Marie Sk{\l}odowska-Curie grant
agreement No 801459 -FP-RESOMUS (to KT). \\

\section*{Supporting Information}
The supporting material includes Table S1 to
S6 and Figures S1 to
S4.

\section*{Data Availability}
Relevant data for the present study are available at
\url{https://github.com/MMunibas/DES2}.

\bibliography{refs}

\providecommand{\latin}[1]{#1}
\makeatletter
\providecommand{\doi}
  {\begingroup\let\do\@makeother\dospecials
  \catcode`\{=1 \catcode`\}=2 \doi@aux}
\providecommand{\doi@aux}[1]{\endgroup\texttt{#1}}
\makeatother
\providecommand*\mcitethebibliography{\thebibliography}
\csname @ifundefined\endcsname{endmcitethebibliography}
  {\let\endmcitethebibliography\endthebibliography}{}
\begin{mcitethebibliography}{85}
\providecommand*\natexlab[1]{#1}
\providecommand*\mciteSetBstSublistMode[1]{}
\providecommand*\mciteSetBstMaxWidthForm[2]{}
\providecommand*\mciteBstWouldAddEndPuncttrue
  {\def\EndOfBibitem{\unskip.}}
\providecommand*\mciteBstWouldAddEndPunctfalse
  {\let\EndOfBibitem\relax}
\providecommand*\mciteSetBstMidEndSepPunct[3]{}
\providecommand*\mciteSetBstSublistLabelBeginEnd[3]{}
\providecommand*\EndOfBibitem{}
\mciteSetBstSublistMode{f}
\mciteSetBstMaxWidthForm{subitem}{(\alph{mcitesubitemcount})}
\mciteSetBstSublistLabelBeginEnd
  {\mcitemaxwidthsubitemform\space}
  {\relax}
  {\relax}

\bibitem[Helbing and Hamm(2023)Helbing, and Hamm]{hamm:2023}
Helbing,~J.; Hamm,~P. Versatile femtosecond laser synchronization for
  multiple-timescale transient infrared spectroscopy. \emph{J. Phys. Chem. A}
  \textbf{2023}, \emph{127}, 6347--6356\relax
\mciteBstWouldAddEndPuncttrue
\mciteSetBstMidEndSepPunct{\mcitedefaultmidpunct}
{\mcitedefaultendpunct}{\mcitedefaultseppunct}\relax
\EndOfBibitem
\bibitem[Piana \latin{et~al.}(2014)Piana, Klepeis, and Shaw]{shaw:2014}
Piana,~S.; Klepeis,~J.~L.; Shaw,~D.~E. Assessing the Accuracy of Physical
  Models Used in Protein-folding Simulations: Quantitative Evidence from Long
  Molecular Dynamics Simulations. \emph{Curr. Op. Struct. Biol.} \textbf{2014},
  \emph{24}, 98--105\relax
\mciteBstWouldAddEndPuncttrue
\mciteSetBstMidEndSepPunct{\mcitedefaultmidpunct}
{\mcitedefaultendpunct}{\mcitedefaultseppunct}\relax
\EndOfBibitem
\bibitem[Simonson \latin{et~al.}(2002)Simonson, Archontis, and
  Karplus]{simonson:2002}
Simonson,~T.; Archontis,~G.; Karplus,~M. Free Energy Simulations Come of Age:
  Protein- Ligand Recognition. \emph{Acc. Chem. Res.} \textbf{2002}, \emph{35},
  430--437\relax
\mciteBstWouldAddEndPuncttrue
\mciteSetBstMidEndSepPunct{\mcitedefaultmidpunct}
{\mcitedefaultendpunct}{\mcitedefaultseppunct}\relax
\EndOfBibitem
\bibitem[Feig and Sugita(2019)Feig, and Sugita]{feig:2019}
Feig,~M.; Sugita,~Y. Whole-cell Models and Simulations in Molecular Detail.
  \emph{Ann. Rev. Cell Devel. Biol.} \textbf{2019}, \emph{35}, 191--211\relax
\mciteBstWouldAddEndPuncttrue
\mciteSetBstMidEndSepPunct{\mcitedefaultmidpunct}
{\mcitedefaultendpunct}{\mcitedefaultseppunct}\relax
\EndOfBibitem
\bibitem[Koner \latin{et~al.}(2020)Koner, Salehi, Mondal, and
  Meuwly]{MM.jcp:2020}
Koner,~D.; Salehi,~S.~M.; Mondal,~P.; Meuwly,~M. Non-conventional Force Fields
  for Applications in Spectroscopy and Chemical Reaction Dynamics. \emph{J.
  Chem. Phys.} \textbf{2020}, \emph{153}, 010901\relax
\mciteBstWouldAddEndPuncttrue
\mciteSetBstMidEndSepPunct{\mcitedefaultmidpunct}
{\mcitedefaultendpunct}{\mcitedefaultseppunct}\relax
\EndOfBibitem
\bibitem[Nibbering \latin{et~al.}(2005)Nibbering, Fidder, and
  Pines]{pines:2005}
Nibbering,~E.~T.; Fidder,~H.; Pines,~E. Ultrafast Chemistry: Using
  Time-resolved Vibrational Spectroscopy for Interrogation of Structural
  Dynamics. \emph{Ann. Rev. Phys. Chem.} \textbf{2005}, \emph{56},
  337--367\relax
\mciteBstWouldAddEndPuncttrue
\mciteSetBstMidEndSepPunct{\mcitedefaultmidpunct}
{\mcitedefaultendpunct}{\mcitedefaultseppunct}\relax
\EndOfBibitem
\bibitem[Hamm and Zanni(2011)Hamm, and Zanni]{2dir:2011}
Hamm,~P.; Zanni,~M. \emph{Concepts and Methods of 2D Infrared Spectroscopy};
  Cambridge University Press: Cambridge, 2011\relax
\mciteBstWouldAddEndPuncttrue
\mciteSetBstMidEndSepPunct{\mcitedefaultmidpunct}
{\mcitedefaultendpunct}{\mcitedefaultseppunct}\relax
\EndOfBibitem
\bibitem[Lee \latin{et~al.}(2013)Lee, Carr, G{\"o}llner, Hamm, and
  Meuwly]{MM.cn:2013}
Lee,~M.~W.; Carr,~J.~K.; G{\"o}llner,~M.; Hamm,~P.; Meuwly,~M. 2D IR spectra of
  cyanide in water investigated by molecular dynamics simulations. \emph{J.
  Chem. Phys.} \textbf{2013}, \emph{139}, 054506\relax
\mciteBstWouldAddEndPuncttrue
\mciteSetBstMidEndSepPunct{\mcitedefaultmidpunct}
{\mcitedefaultendpunct}{\mcitedefaultseppunct}\relax
\EndOfBibitem
\bibitem[T\"opfer \latin{et~al.}(2022)T\"opfer, Pasti, Das, Salehi,
  Vazquez-Salazar, Rohrbach, Feurer, Hamm, and Meuwly]{MM.eutectic:2022}
T\"opfer,~K.; Pasti,~A.; Das,~A.; Salehi,~S.~M.; Vazquez-Salazar,~L.~I.;
  Rohrbach,~D.; Feurer,~T.; Hamm,~P.; Meuwly,~M. Structure, Organization, and
  Heterogeneity of Water-Containing Deep Eutectic Solvents. \emph{J. Am. Chem.
  Soc.} \textbf{2022}, \emph{144}, 14170--14180\relax
\mciteBstWouldAddEndPuncttrue
\mciteSetBstMidEndSepPunct{\mcitedefaultmidpunct}
{\mcitedefaultendpunct}{\mcitedefaultseppunct}\relax
\EndOfBibitem
\bibitem[Mackerell(2004)]{mackerell2004}
Mackerell,~A.~D. Empirical Force Fields for Biological Macromolecules: Overview
  and Issues. \emph{J. Comp. Chem.} \textbf{2004}, \emph{25}, 1584--1604\relax
\mciteBstWouldAddEndPuncttrue
\mciteSetBstMidEndSepPunct{\mcitedefaultmidpunct}
{\mcitedefaultendpunct}{\mcitedefaultseppunct}\relax
\EndOfBibitem
\bibitem[Salehi \latin{et~al.}(2019)Salehi, Koner, and Meuwly]{MM.n3:2019}
Salehi,~S.~M.; Koner,~D.; Meuwly,~M. Vibrational Spectroscopy of N$_3^-$ in the
  Gas and Condensed Phase. \emph{J. Phys. Chem. B} \textbf{2019}, \emph{123},
  3282--3290\relax
\mciteBstWouldAddEndPuncttrue
\mciteSetBstMidEndSepPunct{\mcitedefaultmidpunct}
{\mcitedefaultendpunct}{\mcitedefaultseppunct}\relax
\EndOfBibitem
\bibitem[Koner and Meuwly(2020)Koner, and Meuwly]{MM.rkhs:2020}
Koner,~D.; Meuwly,~M. Permutationally invariant, reproducing kernel-based
  potential energy surfaces for polyatomic molecules: From formaldehyde to
  acetone. \emph{J. Chem. Theo. Comp.} \textbf{2020}, \emph{16},
  5474--5484\relax
\mciteBstWouldAddEndPuncttrue
\mciteSetBstMidEndSepPunct{\mcitedefaultmidpunct}
{\mcitedefaultendpunct}{\mcitedefaultseppunct}\relax
\EndOfBibitem
\bibitem[Nandi \latin{et~al.}(2019)Nandi, Qu, and Bowman]{nandi:2019}
Nandi,~A.; Qu,~C.; Bowman,~J.~M. Full and fragmented permutationally invariant
  polynomial potential energy surfaces for trans and cis N-methyl acetamide and
  isomerization saddle points. \emph{J. Chem. Phys.} \textbf{2019},
  \emph{151}\relax
\mciteBstWouldAddEndPuncttrue
\mciteSetBstMidEndSepPunct{\mcitedefaultmidpunct}
{\mcitedefaultendpunct}{\mcitedefaultseppunct}\relax
\EndOfBibitem
\bibitem[Li \latin{et~al.}(2014)Li, Carter, Bowman, Dawes, Xie, and
  Guo]{li:2014}
Li,~J.; Carter,~S.; Bowman,~J.~M.; Dawes,~R.; Xie,~D.; Guo,~H. High-level,
  first-principles, full-dimensional quantum calculation of the ro-vibrational
  spectrum of the simplest criegee intermediate (CH2OO). \emph{J. Phys. Chem.
  Lett.} \textbf{2014}, \emph{5}, 2364--2369\relax
\mciteBstWouldAddEndPuncttrue
\mciteSetBstMidEndSepPunct{\mcitedefaultmidpunct}
{\mcitedefaultendpunct}{\mcitedefaultseppunct}\relax
\EndOfBibitem
\bibitem[Stone(2013)]{Stone2013}
Stone,~A. \emph{The Theory of Intermolecular Forces}; Oxford University Press:
  Cambridge, 2013\relax
\mciteBstWouldAddEndPuncttrue
\mciteSetBstMidEndSepPunct{\mcitedefaultmidpunct}
{\mcitedefaultendpunct}{\mcitedefaultseppunct}\relax
\EndOfBibitem
\bibitem[Handley \latin{et~al.}(2009)Handley, Hawe, Kell, and
  Popelier]{Handley2009}
Handley,~C.~M.; Hawe,~G.~I.; Kell,~D.~B.; Popelier,~P. L.~A. Optimal
  Construction of a Fast and Accurate Polarisable Water Potential Based on
  Multipole Moments Trained by Machine Learning. \emph{Phys. Chem. Chem. Phys.}
  \textbf{2009}, \emph{11}, 6365\relax
\mciteBstWouldAddEndPuncttrue
\mciteSetBstMidEndSepPunct{\mcitedefaultmidpunct}
{\mcitedefaultendpunct}{\mcitedefaultseppunct}\relax
\EndOfBibitem
\bibitem[Bereau \latin{et~al.}(2013)Bereau, Kramer, and Meuwly]{MM.mtp:2013}
Bereau,~T.; Kramer,~C.; Meuwly,~M. Leveraging Symmetries of Static Atomic
  Multipole Electrostatics in Molecular Dynamics Simulations. \emph{J. Chem.
  Theo. Comput.} \textbf{2013}, \emph{9}, 5450--5459\relax
\mciteBstWouldAddEndPuncttrue
\mciteSetBstMidEndSepPunct{\mcitedefaultmidpunct}
{\mcitedefaultendpunct}{\mcitedefaultseppunct}\relax
\EndOfBibitem
\bibitem[Devereux \latin{et~al.}(2014)Devereux, Raghunathan, Fedorov, and
  Meuwly]{Devereux2014}
Devereux,~M.; Raghunathan,~S.; Fedorov,~D.~G.; Meuwly,~M. A Novel,
  Computationally Efficient Multipolar Model Employing Distributed Charges for
  Molecular Dynamics Simulations. \emph{J. Chem. Theo. Comp.} \textbf{2014},
  \emph{10}, 4229\relax
\mciteBstWouldAddEndPuncttrue
\mciteSetBstMidEndSepPunct{\mcitedefaultmidpunct}
{\mcitedefaultendpunct}{\mcitedefaultseppunct}\relax
\EndOfBibitem
\bibitem[Bereau and Meuwly(2016)Bereau, and Meuwly]{bereau:2016}
Bereau,~T.; Meuwly,~M. \emph{Many-Body Effects and Electrostatics in
  Biomolecules}; Jenny Stanford Publishing, 2016; pp 251--286\relax
\mciteBstWouldAddEndPuncttrue
\mciteSetBstMidEndSepPunct{\mcitedefaultmidpunct}
{\mcitedefaultendpunct}{\mcitedefaultseppunct}\relax
\EndOfBibitem
\bibitem[Jing \latin{et~al.}(2019)Jing, Liu, Cheng, Qi, Walker, Piquemal, and
  Ren]{ren:2019}
Jing,~Z.; Liu,~C.; Cheng,~S.~Y.; Qi,~R.; Walker,~B.~D.; Piquemal,~J.-P.;
  Ren,~P. Polarizable Force Fields for Biomolecular Simulations: Recent
  Advances and Applications. \emph{Ann. Rev. Biophys.} \textbf{2019},
  \emph{48}, 371--394\relax
\mciteBstWouldAddEndPuncttrue
\mciteSetBstMidEndSepPunct{\mcitedefaultmidpunct}
{\mcitedefaultendpunct}{\mcitedefaultseppunct}\relax
\EndOfBibitem
\bibitem[Halgren(1992)]{halgren:1992}
Halgren,~T.~A. The Representation of Van Der Waals (vdw) Interactions in
  Molecular Mechanics Force Fields: Potential Form, Combination Rules, and Vdw
  Parameters. \emph{J. Am. Chem. Soc.} \textbf{1992}, \emph{114},
  7827--7843\relax
\mciteBstWouldAddEndPuncttrue
\mciteSetBstMidEndSepPunct{\mcitedefaultmidpunct}
{\mcitedefaultendpunct}{\mcitedefaultseppunct}\relax
\EndOfBibitem
\bibitem[Bzowski \latin{et~al.}(1988)Bzowski, Mason, and Kestin]{mason:1988}
Bzowski,~J.; Mason,~E.; Kestin,~J. On Combination Rules for Molecular Van Der
  Waals Potential-well Parameters. \emph{Int. J. Therm..} \textbf{1988},
  \emph{9}, 131--143\relax
\mciteBstWouldAddEndPuncttrue
\mciteSetBstMidEndSepPunct{\mcitedefaultmidpunct}
{\mcitedefaultendpunct}{\mcitedefaultseppunct}\relax
\EndOfBibitem
\bibitem[Delhommelle and Milli{\'e}(2001)Delhommelle, and
  Milli{\'e}]{millie:2001}
Delhommelle,~J.; Milli{\'e},~P. Inadequacy of the Lorentz-berthelot Combining
  Rules for Accurate Predictions of Equilibrium Properties by Molecular
  Simulation. \emph{Mol. Phys.} \textbf{2001}, \emph{99}, 619--625\relax
\mciteBstWouldAddEndPuncttrue
\mciteSetBstMidEndSepPunct{\mcitedefaultmidpunct}
{\mcitedefaultendpunct}{\mcitedefaultseppunct}\relax
\EndOfBibitem
\bibitem[Abbott \latin{et~al.}(2003)Abbott, Capper, Davies, Rasheed, and
  Tambyrajah]{abbott2003DES}
Abbott,~A.~P.; Capper,~G.; Davies,~D.~L.; Rasheed,~R.~K.; Tambyrajah,~V. Novel
  solvent properties of choline chloride/urea mixtures. \emph{Chem. Commun.}
  \textbf{2003}, 70--71\relax
\mciteBstWouldAddEndPuncttrue
\mciteSetBstMidEndSepPunct{\mcitedefaultmidpunct}
{\mcitedefaultendpunct}{\mcitedefaultseppunct}\relax
\EndOfBibitem
\bibitem[Marcus(2019)]{marcus2019trends}
Marcus,~Y. \emph{Deep Eutectic Solvents}; Springer, 2019; pp 185--191\relax
\mciteBstWouldAddEndPuncttrue
\mciteSetBstMidEndSepPunct{\mcitedefaultmidpunct}
{\mcitedefaultendpunct}{\mcitedefaultseppunct}\relax
\EndOfBibitem
\bibitem[Martins \latin{et~al.}(2019)Martins, Pinho, and
  Coutinho]{martins2019defdes}
Martins,~M.~A.; Pinho,~S.~P.; Coutinho,~J.~A. Insights into the nature of
  eutectic and deep eutectic mixtures. \emph{Journal of Solution Chemistry}
  \textbf{2019}, \emph{48}, 962--982\relax
\mciteBstWouldAddEndPuncttrue
\mciteSetBstMidEndSepPunct{\mcitedefaultmidpunct}
{\mcitedefaultendpunct}{\mcitedefaultseppunct}\relax
\EndOfBibitem
\bibitem[Smith \latin{et~al.}(2014)Smith, Abbott, and Ryder]{smith:2014}
Smith,~E.~L.; Abbott,~A.~P.; Ryder,~K.~S. Deep eutectic solvents (DESs) and
  their applications. \emph{Chem. Rev.} \textbf{2014}, \emph{114},
  11060--11082\relax
\mciteBstWouldAddEndPuncttrue
\mciteSetBstMidEndSepPunct{\mcitedefaultmidpunct}
{\mcitedefaultendpunct}{\mcitedefaultseppunct}\relax
\EndOfBibitem
\bibitem[Isaac and Kerridge(1988)Isaac, and Kerridge]{isaac:1988}
Isaac,~I.~Y.; Kerridge,~D.~H. Molten acetamide--potassium thiocyanate eutectic:
  spectroscopy of first-row transition metal compounds in a room temperature
  melt. \emph{J. Chem. Soc., Dalton Trans.} \textbf{1988}, 2701--2704\relax
\mciteBstWouldAddEndPuncttrue
\mciteSetBstMidEndSepPunct{\mcitedefaultmidpunct}
{\mcitedefaultendpunct}{\mcitedefaultseppunct}\relax
\EndOfBibitem
\bibitem[Kalita \latin{et~al.}(1998)Kalita, Rohman, and Mahiuddin]{kalita:1998}
Kalita,~G.; Rohman,~N.; Mahiuddin,~S. Viscosity and molar volume of potassium
  thiocyanate+ sodium thiocyanate+ acetamide melt systems. \emph{J. Chem. Eng.
  Data} \textbf{1998}, \emph{43}, 148--151\relax
\mciteBstWouldAddEndPuncttrue
\mciteSetBstMidEndSepPunct{\mcitedefaultmidpunct}
{\mcitedefaultendpunct}{\mcitedefaultseppunct}\relax
\EndOfBibitem
\bibitem[Sakpal \latin{et~al.}(2021)Sakpal, Deshmukh, Chatterjee, Ghosh, and
  Bagchi]{sakpal:2021}
Sakpal,~S.~S.; Deshmukh,~S.~H.; Chatterjee,~S.; Ghosh,~D.; Bagchi,~S.
  Transition of a deep eutectic solution to aqueous solution: A dynamical
  perspective of the dissolved solute. \emph{J. Phys. Chem. Lett.}
  \textbf{2021}, \emph{12}, 8784--8789\relax
\mciteBstWouldAddEndPuncttrue
\mciteSetBstMidEndSepPunct{\mcitedefaultmidpunct}
{\mcitedefaultendpunct}{\mcitedefaultseppunct}\relax
\EndOfBibitem
\bibitem[Ferreira \latin{et~al.}(2016)Ferreira, Voroshylova, Pereira, and
  DS~Cordeiro]{ferreira:2016}
Ferreira,~E.~S.; Voroshylova,~I.~V.; Pereira,~C.~M.; DS~Cordeiro,~M.~N.
  Improved force field model for the deep eutectic solvent ethaline: Reliable
  physicochemical properties. \emph{J. Phys. Chem. B} \textbf{2016},
  \emph{120}, 10124--10137\relax
\mciteBstWouldAddEndPuncttrue
\mciteSetBstMidEndSepPunct{\mcitedefaultmidpunct}
{\mcitedefaultendpunct}{\mcitedefaultseppunct}\relax
\EndOfBibitem
\bibitem[Goloviznina \latin{et~al.}(2022)Goloviznina, Gong, and
  Padua]{padua:2022}
Goloviznina,~K.; Gong,~Z.; Padua,~A.~A. The CL \&Pol polarizable force field
  for the simulation of ionic liquids and eutectic solvents. \emph{Wiley
  Intern. Rev. Comp. Mol. Sci.} \textbf{2022}, \emph{12}, e1572\relax
\mciteBstWouldAddEndPuncttrue
\mciteSetBstMidEndSepPunct{\mcitedefaultmidpunct}
{\mcitedefaultendpunct}{\mcitedefaultseppunct}\relax
\EndOfBibitem
\bibitem[Maglia~de Souza \latin{et~al.}(2021)Maglia~de Souza, Karttunen, and
  Ribeiro]{maglia:2021}
Maglia~de Souza,~R.; Karttunen,~M.; Ribeiro,~M. C.~C. Fine-tuning the
  polarizable CL\&Pol force field for the deep eutectic solvent ethaline.
  \emph{J. Chem. Inf. Model.} \textbf{2021}, \emph{61}, 5938--5947\relax
\mciteBstWouldAddEndPuncttrue
\mciteSetBstMidEndSepPunct{\mcitedefaultmidpunct}
{\mcitedefaultendpunct}{\mcitedefaultseppunct}\relax
\EndOfBibitem
\bibitem[Doherty and Acevedo(2018)Doherty, and Acevedo]{doherty:2018}
Doherty,~B.; Acevedo,~O. OPLS force field for choline chloride-based deep
  eutectic solvents. \emph{J. Phys. Chem. B} \textbf{2018}, \emph{122},
  9982--9993\relax
\mciteBstWouldAddEndPuncttrue
\mciteSetBstMidEndSepPunct{\mcitedefaultmidpunct}
{\mcitedefaultendpunct}{\mcitedefaultseppunct}\relax
\EndOfBibitem
\bibitem[Jeong \latin{et~al.}(2021)Jeong, McDaniel, and Yethiraj]{jeong:2021}
Jeong,~K.-j.; McDaniel,~J.~G.; Yethiraj,~A. Deep eutectic solvents: Molecular
  simulations with a first-principles polarizable force field. \emph{J. Phys.
  Chem. B} \textbf{2021}, \emph{125}, 7177--7186\relax
\mciteBstWouldAddEndPuncttrue
\mciteSetBstMidEndSepPunct{\mcitedefaultmidpunct}
{\mcitedefaultendpunct}{\mcitedefaultseppunct}\relax
\EndOfBibitem
\bibitem[Garc{\'\i}a \latin{et~al.}(2015)Garc{\'\i}a, Atilhan, and
  Aparicio]{garcia:2015}
Garc{\'\i}a,~G.; Atilhan,~M.; Aparicio,~S. The impact of charges in force field
  parameterization for molecular dynamics simulations of deep eutectic
  solvents. \emph{J. Mol. Liq.} \textbf{2015}, \emph{211}, 506--514\relax
\mciteBstWouldAddEndPuncttrue
\mciteSetBstMidEndSepPunct{\mcitedefaultmidpunct}
{\mcitedefaultendpunct}{\mcitedefaultseppunct}\relax
\EndOfBibitem
\bibitem[Zhang \latin{et~al.}(2022)Zhang, Squire, Gurkan, and
  Maginn]{zhang:2022}
Zhang,~Y.; Squire,~H.; Gurkan,~B.; Maginn,~E.~J. Refined classical force field
  for choline chloride and ethylene glycol mixtures over wide composition
  range. \emph{J. Chem. Eng. Data} \textbf{2022}, \emph{67}, 1864--1871\relax
\mciteBstWouldAddEndPuncttrue
\mciteSetBstMidEndSepPunct{\mcitedefaultmidpunct}
{\mcitedefaultendpunct}{\mcitedefaultseppunct}\relax
\EndOfBibitem
\bibitem[Velez and Acevedo(2022)Velez, and Acevedo]{velez:2022}
Velez,~C.; Acevedo,~O. Simulation of deep eutectic solvents: Progress to
  promises. \emph{Wiley Intern. Rev. Comp. Mol. Sci.} \textbf{2022}, \emph{12},
  e1598\relax
\mciteBstWouldAddEndPuncttrue
\mciteSetBstMidEndSepPunct{\mcitedefaultmidpunct}
{\mcitedefaultendpunct}{\mcitedefaultseppunct}\relax
\EndOfBibitem
\bibitem[Brooks \latin{et~al.}(2009)Brooks, Brooks~III, MacKerell~Jr., Nilsson,
  Petrella, Roux, Won, Archontis, Bartels, Boresch, Caflisch, Caves, Cui,
  Dinner, Feig, Fischer, Gao, Hodoscek, Im, Kuczera, Lazaridis, Ma,
  Ovchinnikov, Paci, Pastor, Post, Schaefer, Tidor, Venable, Woodcock, Wu,
  Yang, York, and Karplus]{Charmm-Brooks-2009}
Brooks,~B.~R.; Brooks~III,~C.~L.; MacKerell~Jr.,~A.~D.; Nilsson,~L.;
  Petrella,~R.~J.; Roux,~B.; Won,~Y.; Archontis,~G.; Bartels,~C.; Boresch,~S.
  \latin{et~al.}  CHARMM: The Biomolecular Simulation Program. \emph{J. Comp.
  Chem.} \textbf{2009}, \emph{30}, 1545--1614\relax
\mciteBstWouldAddEndPuncttrue
\mciteSetBstMidEndSepPunct{\mcitedefaultmidpunct}
{\mcitedefaultendpunct}{\mcitedefaultseppunct}\relax
\EndOfBibitem
\bibitem[Ryckaert \latin{et~al.}(1977)Ryckaert, Ciccotti, and
  Berendsen]{shake77}
Ryckaert,~J.-P.; Ciccotti,~G.; Berendsen,~H. J.~C. Numerical integration of the
  cartesian equations of motion of a system with constraints: molecular
  dynamics of n-alkanes. \emph{J. Chem. Phys.} \textbf{1977}, \emph{23},
  327--341\relax
\mciteBstWouldAddEndPuncttrue
\mciteSetBstMidEndSepPunct{\mcitedefaultmidpunct}
{\mcitedefaultendpunct}{\mcitedefaultseppunct}\relax
\EndOfBibitem
\bibitem[Darden \latin{et~al.}(1993)Darden, York, and Pedersen]{Darden1993}
Darden,~T.; York,~D.; Pedersen,~L. Particle mesh Ewald: An N*log(N) method for
  Ewald sums in large systems. \emph{J. Chem. Phys.} \textbf{1993}, \emph{98},
  10089--10092\relax
\mciteBstWouldAddEndPuncttrue
\mciteSetBstMidEndSepPunct{\mcitedefaultmidpunct}
{\mcitedefaultendpunct}{\mcitedefaultseppunct}\relax
\EndOfBibitem
\bibitem[Mart\'{i}nez \latin{et~al.}(2009)Mart\'{i}nez, Andrade, Birgin, and
  Mart\'{i}nez]{martinez:2009}
Mart\'{i}nez,~L.; Andrade,~R.; Birgin,~E.~G.; Mart\'{i}nez,~J.~M. Packmol: A
  Package for Building Initial Configurations for Molecular Dynamics
  Simulations. \emph{J. Comp. Chem.} \textbf{2009}, \emph{30}, 2157--2164\relax
\mciteBstWouldAddEndPuncttrue
\mciteSetBstMidEndSepPunct{\mcitedefaultmidpunct}
{\mcitedefaultendpunct}{\mcitedefaultseppunct}\relax
\EndOfBibitem
\bibitem[Hoover(1985)]{Hoover1985}
Hoover,~W.~G. Canonical dynamics: Equilibrium phase-space distributions.
  \emph{Phys. Rev. A} \textbf{1985}, \emph{31}, 1695--1697\relax
\mciteBstWouldAddEndPuncttrue
\mciteSetBstMidEndSepPunct{\mcitedefaultmidpunct}
{\mcitedefaultendpunct}{\mcitedefaultseppunct}\relax
\EndOfBibitem
\bibitem[Feller \latin{et~al.}(1995)Feller, Zhang, Pastor, and
  Brooks]{Brooks1995}
Feller,~S.~E.; Zhang,~Y.; Pastor,~R.~W.; Brooks,~B.~R. {Constant pressure
  molecular dynamics simulation: The Langevin piston method}. \emph{J. Chem.
  Phys.} \textbf{1995}, \emph{103}, 4613--4621\relax
\mciteBstWouldAddEndPuncttrue
\mciteSetBstMidEndSepPunct{\mcitedefaultmidpunct}
{\mcitedefaultendpunct}{\mcitedefaultseppunct}\relax
\EndOfBibitem
\bibitem[Vanommeslaeghe \latin{et~al.}(2010)Vanommeslaeghe, Hatcher, Acharya,
  Kundu, Zhong, Shim, Darian, Guvench, Lopes, Vorobyov, and
  Mackerell~Jr.]{cgenff}
Vanommeslaeghe,~K.; Hatcher,~E.; Acharya,~C.; Kundu,~S.; Zhong,~S.; Shim,~J.;
  Darian,~E.; Guvench,~O.; Lopes,~P.; Vorobyov,~I. \latin{et~al.}  CHARMM
  general force field: A force field for drug-like molecules compatible with
  the CHARMM all-atom additive biological force fields. \emph{J. Comp. Chem.}
  \textbf{2010}, \emph{31}, 671--690\relax
\mciteBstWouldAddEndPuncttrue
\mciteSetBstMidEndSepPunct{\mcitedefaultmidpunct}
{\mcitedefaultendpunct}{\mcitedefaultseppunct}\relax
\EndOfBibitem
\bibitem[Jorgensen \latin{et~al.}(1983)Jorgensen, Chandrasekhar, Madura, Impey,
  and Klein]{TIP3P-Jorgensen-1983}
Jorgensen,~W.~L.; Chandrasekhar,~J.; Madura,~J.~D.; Impey,~R.~W.; Klein,~M.~L.
  Comparison of Simple Potential Functions for Simulating Liquid Water.
  \emph{J. Chem. Phys.} \textbf{1983}, \emph{79}, 926--935\relax
\mciteBstWouldAddEndPuncttrue
\mciteSetBstMidEndSepPunct{\mcitedefaultmidpunct}
{\mcitedefaultendpunct}{\mcitedefaultseppunct}\relax
\EndOfBibitem
\bibitem[Bian \latin{et~al.}(2013)Bian, Chen, Zhang, Li, Wen, Zhuang, and
  Zheng]{bian:2013}
Bian,~H.; Chen,~H.; Zhang,~Q.; Li,~J.; Wen,~X.; Zhuang,~W.; Zheng,~J. Cation
  effects on rotational dynamics of anions and water molecules in alkali
  (Li$^{+}$, Na$^{+}$, K$^{+}$, Cs$^{+}$) thiocyanate (SCN$^{-}$) aqueous
  solutions. \emph{J. Phys. Chem. B} \textbf{2013}, \emph{117},
  7972--7984\relax
\mciteBstWouldAddEndPuncttrue
\mciteSetBstMidEndSepPunct{\mcitedefaultmidpunct}
{\mcitedefaultendpunct}{\mcitedefaultseppunct}\relax
\EndOfBibitem
\bibitem[Tesei \latin{et~al.}(2018)Tesei, Aspelin, and Lund]{lund:2018}
Tesei,~G.; Aspelin,~V.; Lund,~M. Specific Cation Effects on SCN$^{-}$ in Bulk
  Solution and at the Air–Water Interface. \emph{J. Phys. Chem. B}
  \textbf{2018}, \emph{122}, 5094--5105\relax
\mciteBstWouldAddEndPuncttrue
\mciteSetBstMidEndSepPunct{\mcitedefaultmidpunct}
{\mcitedefaultendpunct}{\mcitedefaultseppunct}\relax
\EndOfBibitem
\bibitem[Ho and Rabitz(1996)Ho, and Rabitz]{rabitz:1996}
Ho,~T.; Rabitz,~H. {A general method for constructing multidimensional
  molecular potential energy surfaces from ab-initio calculations}. \emph{J.
  Chem. Phys.} \textbf{1996}, \emph{104}, 2584--2597\relax
\mciteBstWouldAddEndPuncttrue
\mciteSetBstMidEndSepPunct{\mcitedefaultmidpunct}
{\mcitedefaultendpunct}{\mcitedefaultseppunct}\relax
\EndOfBibitem
\bibitem[Unke and Meuwly(2017)Unke, and Meuwly]{MM.rkhs:2017}
Unke,~O.~T.; Meuwly,~M. Toolkit for the construction of reproducing
  kernel-based representations of data: Application to multidimensional
  potential energy surfaces. \emph{J. Chem. Theo. Comp.} \textbf{2017},
  \emph{57}, 1923--1931\relax
\mciteBstWouldAddEndPuncttrue
\mciteSetBstMidEndSepPunct{\mcitedefaultmidpunct}
{\mcitedefaultendpunct}{\mcitedefaultseppunct}\relax
\EndOfBibitem
\bibitem[Ma \latin{et~al.}(2017)Ma, Schwilk, Köppl, and Werner]{werner:2017}
Ma,~Q.; Schwilk,~M.; Köppl,~C.; Werner,~H.-J. Scalable Electron Correlation
  Methods. 4. Parallel Explicitly Correlated Local Coupled Cluster with Pair
  Natural Orbitals (PNO-LCCSD-F12). \emph{J. Chem. Theo. Comp.} \textbf{2017},
  \emph{13}, 4871--4896\relax
\mciteBstWouldAddEndPuncttrue
\mciteSetBstMidEndSepPunct{\mcitedefaultmidpunct}
{\mcitedefaultendpunct}{\mcitedefaultseppunct}\relax
\EndOfBibitem
\bibitem[Werner \latin{et~al.}(2020)Werner, Knowles, Manby, Black, Doll,
  Heßelmann, Kats, Köhn, Korona, Kreplin, Ma, Miller, Mitrushchenkov,
  Peterson, Polyak, Rauhut, and Sibaev]{werner:2020}
Werner,~H.-J.; Knowles,~P.~J.; Manby,~F.~R.; Black,~J.~A.; Doll,~K.;
  Heßelmann,~A.; Kats,~D.; Köhn,~A.; Korona,~T.; Kreplin,~D.~A.
  \latin{et~al.}  The Molpro quantum chemistry package. \emph{J. Chem. Phys.}
  \textbf{2020}, \emph{152}, 144107\relax
\mciteBstWouldAddEndPuncttrue
\mciteSetBstMidEndSepPunct{\mcitedefaultmidpunct}
{\mcitedefaultendpunct}{\mcitedefaultseppunct}\relax
\EndOfBibitem
\bibitem[Boittier \latin{et~al.}(2022)Boittier, Devereux, and
  Meuwly]{MM.fmdcm:2022}
Boittier,~E.~D.; Devereux,~M.; Meuwly,~M. Molecular Dynamics with
  Conformationally Dependent, Distributed Charges. \emph{J. Chem. Theo. Comp.}
  \textbf{2022}, \emph{18}, 7544--7554\relax
\mciteBstWouldAddEndPuncttrue
\mciteSetBstMidEndSepPunct{\mcitedefaultmidpunct}
{\mcitedefaultendpunct}{\mcitedefaultseppunct}\relax
\EndOfBibitem
\bibitem[Frisch \latin{et~al.}(2016)Frisch, Trucks, Schlegel, Scuseria, Robb,
  Cheeseman, Scalmani, Barone, Petersson, Nakatsuji, Li, Caricato, Marenich,
  Bloino, Janesko, Gomperts, Mennucci, Hratchian, Ortiz, Izmaylov, Sonnenberg,
  Williams-Young, Ding, Lipparini, Egidi, Goings, Peng, Petrone, Henderson,
  Ranasinghe, Zakrzewski, Gao, Rega, Zheng, Liang, Hada, Ehara, Toyota, Fukuda,
  Hasegawa, Ishida, Nakajima, Honda, Kitao, Nakai, Vreven, Throssell,
  Montgomery, Peralta, Ogliaro, Bearpark, Heyd, Brothers, Kudin, Staroverov,
  Keith, Kobayashi, Normand, Raghavachari, Rendell, Burant, Iyengar, Tomasi,
  Cossi, Millam, Klene, Adamo, Cammi, Ochterski, Martin, Morokuma, Farkas,
  Foresman, and Fox]{gaussian16}
Frisch,~M.~J.; Trucks,~G.~W.; Schlegel,~H.~B.; Scuseria,~G.~E.; Robb,~M.~A.;
  Cheeseman,~J.~R.; Scalmani,~G.; Barone,~V.; Petersson,~G.~A.; Nakatsuji,~H.
  \latin{et~al.}  Gaussian˜16 {R}evision {C}.01. 2016; Gaussian Inc.
  Wallingford CT\relax
\mciteBstWouldAddEndPuncttrue
\mciteSetBstMidEndSepPunct{\mcitedefaultmidpunct}
{\mcitedefaultendpunct}{\mcitedefaultseppunct}\relax
\EndOfBibitem
\bibitem[Unke \latin{et~al.}(2017)Unke, Devereux, and Meuwly]{MM.mdcm:2017}
Unke,~O.~T.; Devereux,~M.; Meuwly,~M. Minimal distributed charges: Multipolar
  quality at the cost of point charge elect rostatics. \emph{J. Chem. Phys.}
  \textbf{2017}, \emph{147}, 161712\relax
\mciteBstWouldAddEndPuncttrue
\mciteSetBstMidEndSepPunct{\mcitedefaultmidpunct}
{\mcitedefaultendpunct}{\mcitedefaultseppunct}\relax
\EndOfBibitem
\bibitem[Devereux \latin{et~al.}(2020)Devereux, Pezzella, Raghunathan, and
  Meuwly]{MM.mdcm:2020}
Devereux,~M.; Pezzella,~M.; Raghunathan,~S.; Meuwly,~M. Polarizable Multipolar
  Molecular Dynamics Using Distributed Point Charges. \emph{J. Chem. Theo.
  Comp.} \textbf{2020}, \emph{16}, 7267--7280\relax
\mciteBstWouldAddEndPuncttrue
\mciteSetBstMidEndSepPunct{\mcitedefaultmidpunct}
{\mcitedefaultendpunct}{\mcitedefaultseppunct}\relax
\EndOfBibitem
\bibitem[Nocedal and Wright(2006)Nocedal, and Wright]{Nocedal2006Numerical}
Nocedal,~J.; Wright,~S.~J. \emph{Numerical Optimization}, 2nd ed.; Springer
  Series in Operations Research and Financial Engineering; Springer: New York,
  2006\relax
\mciteBstWouldAddEndPuncttrue
\mciteSetBstMidEndSepPunct{\mcitedefaultmidpunct}
{\mcitedefaultendpunct}{\mcitedefaultseppunct}\relax
\EndOfBibitem
\bibitem[Petersen \latin{et~al.}(2005)Petersen, Saykally, Mucha, and
  Jungwirth]{petersen2005scnpol}
Petersen,~P.~B.; Saykally,~R.~J.; Mucha,~M.; Jungwirth,~P. Enhanced
  Concentration of Polarizable Anions at the Liquid Water Surface: SHG
  Spectroscopy and MD Simulations of Sodium Thiocyanide. \emph{J. Phys. Chem.
  B} \textbf{2005}, \emph{109}, 10915--10921\relax
\mciteBstWouldAddEndPuncttrue
\mciteSetBstMidEndSepPunct{\mcitedefaultmidpunct}
{\mcitedefaultendpunct}{\mcitedefaultseppunct}\relax
\EndOfBibitem
\bibitem[Salvador \latin{et~al.}(2003)Salvador, Curtis, Tobias, and
  Jungwirth]{salvador2003pol}
Salvador,~P.; Curtis,~J.~E.; Tobias,~D.~J.; Jungwirth,~P. Polarizability of the
  nitrate anion and its solvation at the air/water interface. \emph{Phys. Chem.
  Chem. Phys.} \textbf{2003}, \emph{5}, 3752--3757\relax
\mciteBstWouldAddEndPuncttrue
\mciteSetBstMidEndSepPunct{\mcitedefaultmidpunct}
{\mcitedefaultendpunct}{\mcitedefaultseppunct}\relax
\EndOfBibitem
\bibitem[Jungwirth and Tobias(2002)Jungwirth, and Tobias]{jungwirth2002pol}
Jungwirth,~P.; Tobias,~D.~J. Chloride Anion on Aqueous Clusters, at the
  Air-Water Interface, and in Liquid Water: Solvent Effects on Cl$^-$
  Polarizability. \emph{J. Phys. Chem. A} \textbf{2002}, \emph{106},
  379--383\relax
\mciteBstWouldAddEndPuncttrue
\mciteSetBstMidEndSepPunct{\mcitedefaultmidpunct}
{\mcitedefaultendpunct}{\mcitedefaultseppunct}\relax
\EndOfBibitem
\bibitem[Hedin \latin{et~al.}(2016)Hedin, El~Hage, and Meuwly]{mm.mtp2:2016}
Hedin,~F.; El~Hage,~K.; Meuwly,~M. A Toolkit to Fit Nonbonded Parameters from
  and for Condensed Phase Simulations. \emph{J. Chem. Inf. Model.}
  \textbf{2016}, \emph{56}, 1479--1489\relax
\mciteBstWouldAddEndPuncttrue
\mciteSetBstMidEndSepPunct{\mcitedefaultmidpunct}
{\mcitedefaultendpunct}{\mcitedefaultseppunct}\relax
\EndOfBibitem
\bibitem[Hedin \latin{et~al.}(2017)Hedin, El~Hage, and Meuwly]{mm.mtp:2017}
Hedin,~F.; El~Hage,~K.; Meuwly,~M. A Toolkit to Fit Nonbonded Parameters from
  and for Condensed Phase Simulations (vol 56, pg 1479, 2016). \emph{J. Chem.
  Inf. Model.} \textbf{2017}, \emph{57}, 102--103\relax
\mciteBstWouldAddEndPuncttrue
\mciteSetBstMidEndSepPunct{\mcitedefaultmidpunct}
{\mcitedefaultendpunct}{\mcitedefaultseppunct}\relax
\EndOfBibitem
\bibitem[Devereux \latin{et~al.}(2024)Devereux, Boittier, and
  Meuwly]{MM.ff:2024}
Devereux,~M.; Boittier,~E.~D.; Meuwly,~M. Systematic improvement of empirical
  energy functions in the era of machine learning. \emph{J. Comp. Chem.}
  \textbf{2024}, \emph{45}, 1899--1913\relax
\mciteBstWouldAddEndPuncttrue
\mciteSetBstMidEndSepPunct{\mcitedefaultmidpunct}
{\mcitedefaultendpunct}{\mcitedefaultseppunct}\relax
\EndOfBibitem
\bibitem[Boys and Bernardi(1970)Boys, and Bernardi]{boys:1970}
Boys,~S.~F.; Bernardi,~F. The calculation of small molecular interactions by
  the differences of separate total energies. Some procedures with reduced
  errors. \emph{Mol. Phys.} \textbf{1970}, \emph{19}, 553--566\relax
\mciteBstWouldAddEndPuncttrue
\mciteSetBstMidEndSepPunct{\mcitedefaultmidpunct}
{\mcitedefaultendpunct}{\mcitedefaultseppunct}\relax
\EndOfBibitem
\bibitem[Straatsma and Berendsen(1988)Straatsma, and Berendsen]{straatsma:1988}
Straatsma,~T.; Berendsen,~H. Free energy of ionic hydration: Analysis of a
  thermodynamic integration technique to evaluate free energy differences by
  molecular dynamics simulations. \emph{J. Chem. Phys.} \textbf{1988},
  \emph{89}, 5876--5886\relax
\mciteBstWouldAddEndPuncttrue
\mciteSetBstMidEndSepPunct{\mcitedefaultmidpunct}
{\mcitedefaultendpunct}{\mcitedefaultseppunct}\relax
\EndOfBibitem
\bibitem[Mackerell~Jr.(2004)]{mackerell:2004}
Mackerell~Jr.,~A.~D. Empirical force fields for biological macromolecules:
  Overview and issues. \emph{J. Comp. Chem.} \textbf{2004}, \emph{25},
  1584--1604\relax
\mciteBstWouldAddEndPuncttrue
\mciteSetBstMidEndSepPunct{\mcitedefaultmidpunct}
{\mcitedefaultendpunct}{\mcitedefaultseppunct}\relax
\EndOfBibitem
\bibitem[Cho \latin{et~al.}(1994)Cho, Fleming, Saito, Ohmine, and
  Stratt]{stratt:1994}
Cho,~M.; Fleming,~G.~R.; Saito,~S.; Ohmine,~I.; Stratt,~R.~M. Instantaneous
  Normal Mode Analysis of Liquid Water. \emph{J. Chem. Phys.} \textbf{1994},
  \emph{100}, 6672--6683\relax
\mciteBstWouldAddEndPuncttrue
\mciteSetBstMidEndSepPunct{\mcitedefaultmidpunct}
{\mcitedefaultendpunct}{\mcitedefaultseppunct}\relax
\EndOfBibitem
\bibitem[Kozi{\'n}ski \latin{et~al.}(2007)Kozi{\'n}ski, Garrett-Roe, and
  Hamm]{kozinski:2007}
Kozi{\'n}ski,~M.; Garrett-Roe,~S.; Hamm,~P. Vibrational spectral diffusion of
  CN$^-$ in water. \emph{Chem. Phys.} \textbf{2007}, \emph{341}, 5--10\relax
\mciteBstWouldAddEndPuncttrue
\mciteSetBstMidEndSepPunct{\mcitedefaultmidpunct}
{\mcitedefaultendpunct}{\mcitedefaultseppunct}\relax
\EndOfBibitem
\bibitem[Virtanen \latin{et~al.}(2020)Virtanen, Gommers, Oliphant, Haberland,
  Reddy, Cournapeau, Burovski, Peterson, Weckesser, Bright, and {\it et
  al.}]{2020SciPy-NMeth}
Virtanen,~P.; Gommers,~R.; Oliphant,~T.~E.; Haberland,~M.; Reddy,~T.;
  Cournapeau,~D.; Burovski,~E.; Peterson,~P.; Weckesser,~W.; Bright,~J.
  \latin{et~al.}  SciPy 1.0: fundamental algorithms for scientific computing in
  Python. \emph{Nat. Meth.} \textbf{2020}, \emph{17}, 261--272\relax
\mciteBstWouldAddEndPuncttrue
\mciteSetBstMidEndSepPunct{\mcitedefaultmidpunct}
{\mcitedefaultendpunct}{\mcitedefaultseppunct}\relax
\EndOfBibitem
\bibitem[Helbing and Hamm(2011)Helbing, and Hamm]{Helbing2011}
Helbing,~J.; Hamm,~P. {Compact implementation of Fourier transform
  two-dimensional IR spectroscopy without phase ambiguity}. \emph{J. Opt. Soc.
  Am. B} \textbf{2011}, \emph{28}, 171\relax
\mciteBstWouldAddEndPuncttrue
\mciteSetBstMidEndSepPunct{\mcitedefaultmidpunct}
{\mcitedefaultendpunct}{\mcitedefaultseppunct}\relax
\EndOfBibitem
\bibitem[Guo \latin{et~al.}(2015)Guo, Pagano, Li, Kohen, and Cheatum]{Guo2015}
Guo,~Q.; Pagano,~P.; Li,~Y.~L.; Kohen,~A.; Cheatum,~C.~M. Line shape analysis
  of two-dimensional infrared spectra. \emph{J. Chem. Phys.} \textbf{2015},
  \emph{142}, 212427\relax
\mciteBstWouldAddEndPuncttrue
\mciteSetBstMidEndSepPunct{\mcitedefaultmidpunct}
{\mcitedefaultendpunct}{\mcitedefaultseppunct}\relax
\EndOfBibitem
\bibitem[Mitchell \latin{et~al.}(1992)Mitchell, Butler, and
  Albright]{albright:1992}
Mitchell,~J.~P.; Butler,~J.~B.; Albright,~J.~G. Measurement of Mutual Diffusion
  Coefficients, Densities, Viscosities, and Osmotic Coefficients for the System
  KSCN-H$_2$O at 25°C. \emph{J Solution Chem} \textbf{1992}, \emph{21},
  1115--1129\relax
\mciteBstWouldAddEndPuncttrue
\mciteSetBstMidEndSepPunct{\mcitedefaultmidpunct}
{\mcitedefaultendpunct}{\mcitedefaultseppunct}\relax
\EndOfBibitem
\bibitem[Marcus(1997)]{marcus:1997}
Marcus,~Y. \emph{Ion Properties}; Ion Properties 1; Taylor \& Francis,
  1997\relax
\mciteBstWouldAddEndPuncttrue
\mciteSetBstMidEndSepPunct{\mcitedefaultmidpunct}
{\mcitedefaultendpunct}{\mcitedefaultseppunct}\relax
\EndOfBibitem
\bibitem[Pearson(1986)]{pearson:1986}
Pearson,~R.~G. Ionization potentials and electron affinities in aqueous
  solution. \emph{J. Am. Chem. Soc.} \textbf{1986}, \emph{108},
  6109--6114\relax
\mciteBstWouldAddEndPuncttrue
\mciteSetBstMidEndSepPunct{\mcitedefaultmidpunct}
{\mcitedefaultendpunct}{\mcitedefaultseppunct}\relax
\EndOfBibitem
\bibitem[Pliego and Riveros(2000)Pliego, and Riveros]{pliego:2000}
Pliego,~J.~R.; Riveros,~J.~M. New values for the absolute solvation free energy
  of univalent ions in aqueous solution. \emph{Chem. Phys. Lett.}
  \textbf{2000}, \emph{332}, 597--602\relax
\mciteBstWouldAddEndPuncttrue
\mciteSetBstMidEndSepPunct{\mcitedefaultmidpunct}
{\mcitedefaultendpunct}{\mcitedefaultseppunct}\relax
\EndOfBibitem
\bibitem[Botti \latin{et~al.}(2009)Botti, Pagnotta, Bruni, and
  Ricci]{botti:2009}
Botti,~A.; Pagnotta,~S.~E.; Bruni,~F.; Ricci,~M.~A. Solvation of KSCN in Water.
  \emph{J. Phys. Chem. B} \textbf{2009}, \emph{113}, 10014--10021\relax
\mciteBstWouldAddEndPuncttrue
\mciteSetBstMidEndSepPunct{\mcitedefaultmidpunct}
{\mcitedefaultendpunct}{\mcitedefaultseppunct}\relax
\EndOfBibitem
\bibitem[Wang \latin{et~al.}(2004)Wang, Wolf, Caldwell, Kollman, and
  Case]{wang:2004}
Wang,~J.; Wolf,~R.~M.; Caldwell,~J.~W.; Kollman,~P.~A.; Case,~D.~A. Development
  and testing of a general amber force field. \emph{J. Comp. Chem.}
  \textbf{2004}, \emph{25}, 1157--1174\relax
\mciteBstWouldAddEndPuncttrue
\mciteSetBstMidEndSepPunct{\mcitedefaultmidpunct}
{\mcitedefaultendpunct}{\mcitedefaultseppunct}\relax
\EndOfBibitem
\bibitem[Perkins \latin{et~al.}(2014)Perkins, Painter, and
  Colina]{perkins:2014}
Perkins,~S.~L.; Painter,~P.; Colina,~C.~M. Experimental and computational
  studies of choline chloride-based deep eutectic solvents. \emph{J. Chem. Eng.
  Data} \textbf{2014}, \emph{59}, 3652--3662\relax
\mciteBstWouldAddEndPuncttrue
\mciteSetBstMidEndSepPunct{\mcitedefaultmidpunct}
{\mcitedefaultendpunct}{\mcitedefaultseppunct}\relax
\EndOfBibitem
\bibitem[Mark and Nilsson(2001)Mark, and Nilsson]{nilsson:2001}
Mark,~P.; Nilsson,~L. Structure and dynamics of the TIP3P, SPC, and SPC/E water
  models at 298 K. \emph{J. Phys. Chem. A} \textbf{2001}, \emph{105},
  9954--9960\relax
\mciteBstWouldAddEndPuncttrue
\mciteSetBstMidEndSepPunct{\mcitedefaultmidpunct}
{\mcitedefaultendpunct}{\mcitedefaultseppunct}\relax
\EndOfBibitem
\bibitem[Kumar and MacKerell~Jr(2024)Kumar, and MacKerell~Jr]{mackerell:2024}
Kumar,~A.; MacKerell~Jr,~A.~D. FFParam-v2. 0: A Comprehensive Tool for CHARMM
  Additive and Drude Polarizable Force-Field Parameter Optimization and
  Validation. \emph{J. Phys. Chem. B} \textbf{2024}, \emph{128},
  4385--4395\relax
\mciteBstWouldAddEndPuncttrue
\mciteSetBstMidEndSepPunct{\mcitedefaultmidpunct}
{\mcitedefaultendpunct}{\mcitedefaultseppunct}\relax
\EndOfBibitem
\bibitem[Zhang \latin{et~al.}(2020)Zhang, Poe, Heroux, Squire, Doherty, Long,
  Dadmun, Gurkan, Tuckerman, and Maginn]{visc_zhang:2020}
Zhang,~Y.; Poe,~D.; Heroux,~L.; Squire,~H.; Doherty,~B.~W.; Long,~Z.;
  Dadmun,~M.; Gurkan,~B.; Tuckerman,~M.~E.; Maginn,~E.~J. Liquid Structure and
  Transport Properties of the Deep Eutectic Solvent Ethaline. \emph{J. Phys.
  Chem. B} \textbf{2020}, \emph{124}, 5251--5264\relax
\mciteBstWouldAddEndPuncttrue
\mciteSetBstMidEndSepPunct{\mcitedefaultmidpunct}
{\mcitedefaultendpunct}{\mcitedefaultseppunct}\relax
\EndOfBibitem
\bibitem[Liu \latin{et~al.}(2013)Liu, Zhao, and Wei]{liu:2013}
Liu,~B.; Zhao,~J.; Wei,~F. Effects of water on the properties of acetamide-KSCN
  eutectic ionic liquids at several temperatures. \emph{J. Mol. Liq.}
  \textbf{2013}, \emph{187}, 309--313\relax
\mciteBstWouldAddEndPuncttrue
\mciteSetBstMidEndSepPunct{\mcitedefaultmidpunct}
{\mcitedefaultendpunct}{\mcitedefaultseppunct}\relax
\EndOfBibitem
\bibitem[Kadaoluwa~Pathirannahalage
  \latin{et~al.}(2021)Kadaoluwa~Pathirannahalage, Meftahi, Elbourne, Weiss,
  McConville, Padua, Winkler, Costa~Gomes, Greaves, Le, \latin{et~al.}
  others]{kadaoluwa:2021}
Kadaoluwa~Pathirannahalage,~S.~P.; Meftahi,~N.; Elbourne,~A.; Weiss,~A.~C.;
  McConville,~C.~F.; Padua,~A.; Winkler,~D.~A.; Costa~Gomes,~M.;
  Greaves,~T.~L.; Le,~T.~C. \latin{et~al.}  Systematic comparison of the
  structural and dynamic properties of commonly used water models for molecular
  dynamics simulations. \emph{J. Chem. Inf. Model.} \textbf{2021}, \emph{61},
  4521--4536\relax
\mciteBstWouldAddEndPuncttrue
\mciteSetBstMidEndSepPunct{\mcitedefaultmidpunct}
{\mcitedefaultendpunct}{\mcitedefaultseppunct}\relax
\EndOfBibitem
\bibitem[Boittier \latin{et~al.}(2024)Boittier, T\"opfer, Devereux, and
  Meuwly]{MM.kmdcm:2024}
Boittier,~E.; T\"opfer,~K.; Devereux,~M.; Meuwly,~M. Kernel-Based Minimal
  Distributed Charges: A Conformationally Dependent ESP-Model for Molecular
  Simulations. \emph{J. Chem. Theo. Comp.} \textbf{2024}, \emph{20},
  8088--8099\relax
\mciteBstWouldAddEndPuncttrue
\mciteSetBstMidEndSepPunct{\mcitedefaultmidpunct}
{\mcitedefaultendpunct}{\mcitedefaultseppunct}\relax
\EndOfBibitem
\end{mcitethebibliography}


\providecommand{\latin}[1]{#1}
\makeatletter
\providecommand{\doi}
  {\begingroup\let\do\@makeother\dospecials
  \catcode`\{=1 \catcode`\}=2 \doi@aux}
\providecommand{\doi@aux}[1]{\endgroup\texttt{#1}}
\makeatother
\providecommand*\mcitethebibliography{\thebibliography}
\csname @ifundefined\endcsname{endmcitethebibliography}
  {\let\endmcitethebibliography\endthebibliography}{}
\begin{mcitethebibliography}{5}
\providecommand*\natexlab[1]{#1}
\providecommand*\mciteSetBstSublistMode[1]{}
\providecommand*\mciteSetBstMaxWidthForm[2]{}
\providecommand*\mciteBstWouldAddEndPuncttrue
  {\def\EndOfBibitem{\unskip.}}
\providecommand*\mciteBstWouldAddEndPunctfalse
  {\let\EndOfBibitem\relax}
\providecommand*\mciteSetBstMidEndSepPunct[3]{}
\providecommand*\mciteSetBstSublistLabelBeginEnd[3]{}
\providecommand*\EndOfBibitem{}
\mciteSetBstSublistMode{f}
\mciteSetBstMaxWidthForm{subitem}{(\alph{mcitesubitemcount})}
\mciteSetBstSublistLabelBeginEnd
  {\mcitemaxwidthsubitemform\space}
  {\relax}
  {\relax}

\bibitem[Bian \latin{et~al.}(2013)Bian, Chen, Zhang, Li, Wen, Zhuang, and
  Zheng]{bian:2013}
Bian,~H.; Chen,~H.; Zhang,~Q.; Li,~J.; Wen,~X.; Zhuang,~W.; Zheng,~J. Cation
  effects on rotational dynamics of anions and water molecules in alkali
  (Li$^{+}$, Na$^{+}$, K$^{+}$, Cs$^{+}$) thiocyanate (SCN$^{-}$) aqueous
  solutions. \emph{J. Phys. Chem. B} \textbf{2013}, \emph{117},
  7972--7984\relax
\mciteBstWouldAddEndPuncttrue
\mciteSetBstMidEndSepPunct{\mcitedefaultmidpunct}
{\mcitedefaultendpunct}{\mcitedefaultseppunct}\relax
\EndOfBibitem
\bibitem[Vanommeslaeghe \latin{et~al.}(2010)Vanommeslaeghe, Hatcher, Acharya,
  Kundu, Zhong, Shim, Darian, Guvench, Lopes, Vorobyov, and
  Mackerell~Jr.]{cgenff}
Vanommeslaeghe,~K.; Hatcher,~E.; Acharya,~C.; Kundu,~S.; Zhong,~S.; Shim,~J.;
  Darian,~E.; Guvench,~O.; Lopes,~P.; Vorobyov,~I. \latin{et~al.}  CHARMM
  general force field: A force field for drug-like molecules compatible with
  the CHARMM all-atom additive biological force fields. \emph{J. Comp. Chem.}
  \textbf{2010}, \emph{31}, 671--690\relax
\mciteBstWouldAddEndPuncttrue
\mciteSetBstMidEndSepPunct{\mcitedefaultmidpunct}
{\mcitedefaultendpunct}{\mcitedefaultseppunct}\relax
\EndOfBibitem
\bibitem[Jorgensen \latin{et~al.}(1983)Jorgensen, Chandrasekhar, Madura, Impey,
  and Klein]{TIP3P-Jorgensen-1983}
Jorgensen,~W.~L.; Chandrasekhar,~J.; Madura,~J.~D.; Impey,~R.~W.; Klein,~M.~L.
  Comparison of Simple Potential Functions for Simulating Liquid Water.
  \emph{J. Chem. Phys.} \textbf{1983}, \emph{79}, 926--935\relax
\mciteBstWouldAddEndPuncttrue
\mciteSetBstMidEndSepPunct{\mcitedefaultmidpunct}
{\mcitedefaultendpunct}{\mcitedefaultseppunct}\relax
\EndOfBibitem
\bibitem[Tesei \latin{et~al.}(2018)Tesei, Aspelin, and Lund]{lund:2018}
Tesei,~G.; Aspelin,~V.; Lund,~M. Specific Cation Effects on SCN$^{-}$ in Bulk
  Solution and at the Air–Water Interface. \emph{J. Phys. Chem. B}
  \textbf{2018}, \emph{122}, 5094--5105\relax
\mciteBstWouldAddEndPuncttrue
\mciteSetBstMidEndSepPunct{\mcitedefaultmidpunct}
{\mcitedefaultendpunct}{\mcitedefaultseppunct}\relax
\EndOfBibitem
\end{mcitethebibliography}

\end{document}


\date{\today}

\subsection{Supplementary Information}

\begin{table}
\caption{Molar fraction and number of molecules (Water/W,
  Acetamide/ACM, K$^+$ and SCN$^{-}$) used on the MD simulations for
  each component of the systems for given W:ACM mixing
  ratio.}
\label{sitab_composition}
\begin{tabular}{c||ccc||ccc}
\hline\hline
& \multicolumn{3}{c||}{Molar Fraction} & \multicolumn{3}{c}{Number of Molecules} \\ \hline
W:ACM & \multicolumn{1}{c|}{H$_{2}$O}   & \multicolumn{1}{c|}{Acetamide} & KSCN  & \multicolumn{1}{c|}{H$_{2}$O} & \multicolumn{1}{c|}{Acetamide} & KSCN \\
\hline \hline
100                            & \multicolumn{1}{c|}{0.901} & \multicolumn{1}{c|}{0.000}     & 0.099 & \multicolumn{1}{c|}{685} & \multicolumn{1}{c|}{0}         & 75   \\
90                             & \multicolumn{1}{c|}{0.791} & \multicolumn{1}{c|}{0.092}     & 0.118 & \multicolumn{1}{c|}{506} & \multicolumn{1}{c|}{59}        & 75   \\
80                             & \multicolumn{1}{c|}{0.686} & \multicolumn{1}{c|}{0.179}     & 0.135 & \multicolumn{1}{c|}{381} & \multicolumn{1}{c|}{100}       & 75   \\
70                             & \multicolumn{1}{c|}{0.586} & \multicolumn{1}{c|}{0.263}     & 0.152 & \multicolumn{1}{c|}{290} & \multicolumn{1}{c|}{130}       & 75   \\
50                             & \multicolumn{1}{c|}{0.399} & \multicolumn{1}{c|}{0.418}     & 0.183 & \multicolumn{1}{c|}{164} & \multicolumn{1}{c|}{171}       & 75   \\
40                             & \multicolumn{1}{c|}{0.312} & \multicolumn{1}{c|}{0.490}     & 0.198 & \multicolumn{1}{c|}{119} & \multicolumn{1}{c|}{186}       & 75   \\
30                             & \multicolumn{1}{c|}{0.229} & \multicolumn{1}{c|}{0.559}     & 0.212 & \multicolumn{1}{c|}{81}  & \multicolumn{1}{c|}{198}       & 75   \\
20                             & \multicolumn{1}{c|}{0.149} & \multicolumn{1}{c|}{0.626}     & 0.225 & \multicolumn{1}{c|}{50}  & \multicolumn{1}{c|}{209}       & 75   \\
0                              & \multicolumn{1}{c|}{0.000} & \multicolumn{1}{c|}{0.750}     & 0.250 & \multicolumn{1}{c|}{0}   & \multicolumn{1}{c|}{225}       & 75   \\
\hline
\hline
\end{tabular}
\end{table}

\begin{figure}
  \centering
  \includegraphics[width=0.75\textwidth]{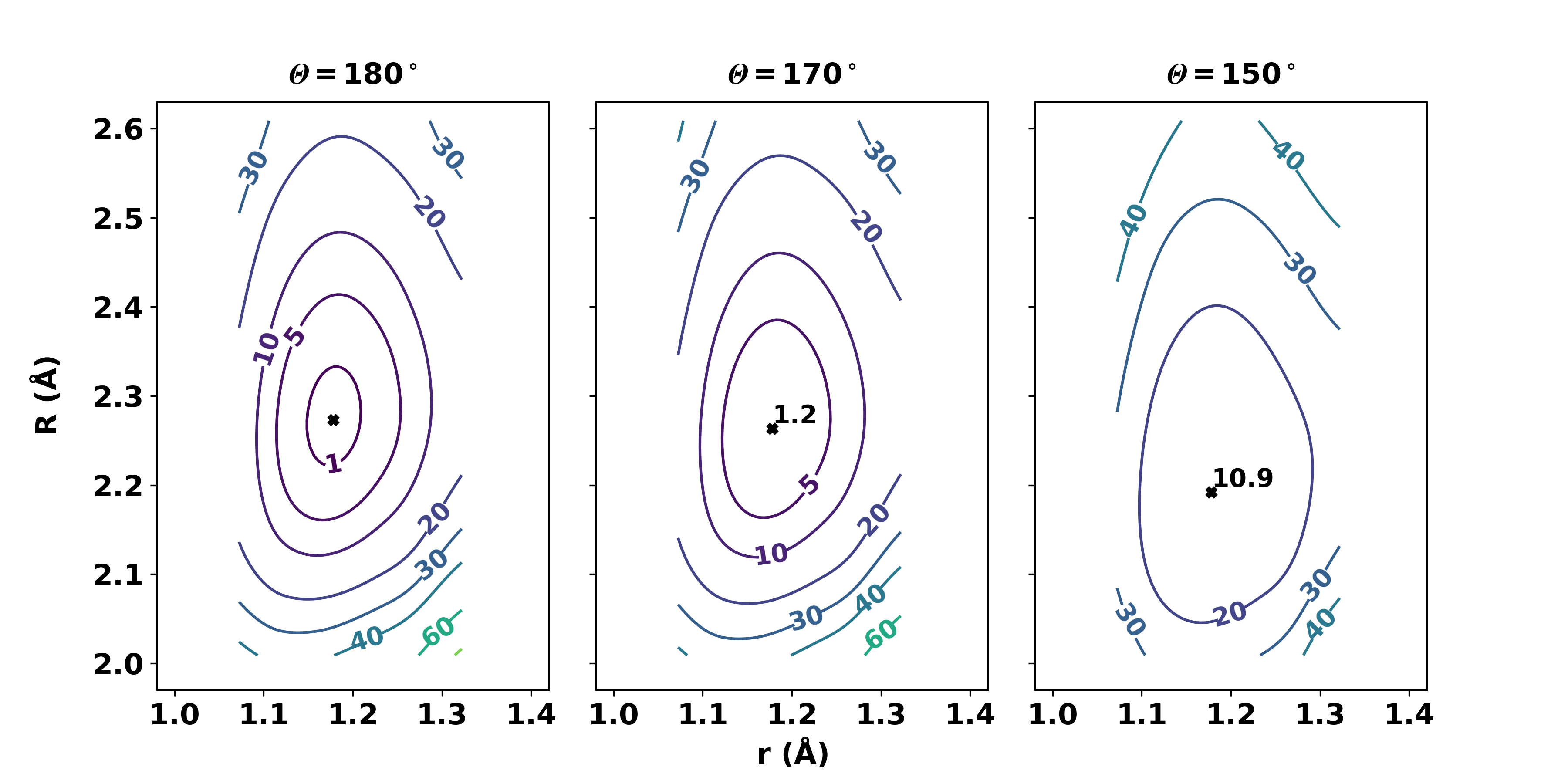}
\caption{2-dimensional PES cuts of the bonding potential of SCN$^-$
  reproduced by an RKHS model for a selection of bond angles
  $\theta$. The black cross marks the potential minimum for each bond
  angle $\theta$ and the zero of energy is at $\theta=180^\circ$.  All
  energies in kcal/mol.}
\label{sifig_scn_rkhs}
\end{figure}

\begin{table}
\caption{Bonded and non-bonded parameters for Model {\bf M0}.
  Nonbonded LJ parameters for SCN$^-$ are adopted from
  Ref. \citenum{bian:2013} with $r_\mathrm{min}$ scaled by $f=1.1$.}
\label{sitab_params_m0}
\begin{tabular}{c|ccc}
\hline\hline
\textbf{Residues} & \multicolumn{3}{c}{Parameters} \\
\hline \hline
\textbf{Acetamide} & \multicolumn{3}{c}{CGenFF\cite{cgenff}} \\
\hline \hline
\textbf{Water} & \multicolumn{3}{c}{TIP3P\cite{TIP3P-Jorgensen-1983}} \\
\hline \hline
\textbf{K$^+$} & \multicolumn{3}{c}{} \\
\hline
Nonbonded\cite{bian:2013} & $\epsilon$ (kcal/mol)& $r_\mathrm{min}$ (\AA) & $q$ (e)\\
K$^+$ & $0.1004$ & $3.7378$ & $+1$ \\
\hline \hline 
\textbf{SCN$^-$} & \multicolumn{3}{c}{} \\
\hline
Bonded & \multicolumn{3}{c}{RKHS (see $^a$)} \\
\hline 
Nonbonded & $\epsilon$ (kcal/mol)& $r_\mathrm{min}$ (\AA) & $q$ (e) \\
S & $0.3639$ & $4.3462\,(=1.1\cdot3.9510)$ & $-0.183$ \\
C & $0.0741$ & $4.0870\,(=1.1\cdot3.7154)$ & $-0.455$ \\
N & $0.1016$ & $4.1362\,(=1.1\cdot3.7602)$ & $-0.362$ \\
\hline 
Atomic Multipoles & S & C & N \\
$Q_{00}$  & $-0.183$ & $-0.362$ & $-0.455$ \\
$Q_{10}$  & $1.179$ & $0.163$ & $0.319$ \\
$Q_{11c}$ & $0.0$ & $0.0$ & $0.0$ \\
$Q_{11s}$ & $0.0$ & $0.0$ & $0.0$ \\
$Q_{20}$  & $-1.310$ & $-0.929$ & $-3.114$ \\
$Q_{21c}$ & $0.0$ & $0.0$ & $0.0$ \\
$Q_{21s}$ & $0.0$ & $0.0$ & $0.0$ \\
$Q_{22c}$ & $0.0$ & $0.0$ & $0.0$ \\
$Q_{22s}$ & $0.0$ & $0.0$ & $0.0$ \\
\hline \hline
\end{tabular}
$^a$\url{https://github.com/MMunibas/DES2/blob/main/M0/source/rkhs_SCN_rRz.csv}
\end{table}

\begin{table}
\caption{Bonded and non-bonded parameters for Model {\bf M1}.
  Nonbonded LJ parameters for SCN$^-$ are adopted from
  Ref. \citenum{lund:2018} with $r_\mathrm{min}$ scaled by $f =
  0.96$.}
\label{sitab_params_m1}
\begin{tabular}{c|ccc}
\hline\hline
\textbf{Residues} & \multicolumn{3}{c}{Parameters} \\
\hline \hline
\textbf{Acetamide} & \multicolumn{3}{c}{CGenFF\cite{cgenff}} \\
\hline \hline
\textbf{Water} & \multicolumn{3}{c}{TIP3P\cite{TIP3P-Jorgensen-1983}} \\
\hline \hline
\textbf{K$^+$} & \multicolumn{3}{c}{} \\
\hline
Nonbonded\cite{lund:2018} & $\epsilon$ (kcal/mol)& $r_\mathrm{min}$ (\AA) & $q$ (e)\\
K$^+$ & $0.2032$ & $4.5234$ & $+1$ \\
\hline \hline 
\textbf{SCN$^-$} & \multicolumn{3}{c}{} \\
\hline
Bonded & \multicolumn{3}{c}{RKHS (see $^a$)} \\
\hline 
Nonbonded & $\epsilon$ (kcal/mol)& $r_\mathrm{min}$ (\AA) & $q$ (e) \\
S & $0.3640$ & $4.3540\,(=0.96\cdot4.5354)$ & $-$ \\
C & $0.0741$ & $4.0260\,(=0.96\cdot4.1938)$ & $-$ \\
N & $0.1016$ & $3.5836\,(=0.96\cdot3.7330)$ & $-$ \\
\hline 
Electrostatic Model &  \multicolumn{3}{c}{fMDCM (see $^b$)}\\
\hline \hline
\end{tabular}
$^a$\url{https://github.com/MMunibas/DES2/blob/main/M1/source/rkhs_SCN_rRz.csv}
$^b$\url{https://github.com/MMunibas/DES2/blob/main/M1/source/scn_fluc.dcm}
\end{table}

\begin{table}
\caption{Bonded and non-bonded parameters for Model {\bf M2}.
  Nonbonded LJ parameters for SCN$^-$ are adopted from cluster
  interaction energy fit (with $r_\mathrm{min}$ scaled by $f = 1.0$).}
\label{sitab_params_m2}
\begin{tabular}{c|ccc}
\hline\hline
\textbf{Residues} & \multicolumn{3}{c}{Parameters} \\
\hline \hline
\textbf{Acetamide} & \multicolumn{3}{c}{CGenFF\cite{cgenff}} \\
\hline \hline
\textbf{Water} & \multicolumn{3}{c}{TIP3P\cite{TIP3P-Jorgensen-1983}} \\
\hline \hline
\textbf{K$^+$} & \multicolumn{3}{c}{} \\
\hline
Nonbonded\cite{bian:2013} & $\epsilon$ (kcal/mol)& $r_\mathrm{min}$ (\AA) & $q$ (e)\\
K$^+$ & $0.1004$ & $3.7378$ & $+1$ \\
\hline \hline 
\textbf{SCN$^-$} & \multicolumn{3}{c}{} \\
\hline
Bonded & \multicolumn{3}{c}{RKHS (see $^a$)} \\
\hline 
Nonbonded & $\epsilon$ (kcal/mol)& $r_\mathrm{min}$ (\AA) & $q$ (e) \\
S & $0.1836$ & $4.8558$ & $-$ \\
C & $0.0001$ & $3.7868$ & $-$ \\
N & $0.0223$ & $4.5720$ & $-$ \\
\hline 
Electrostatic Model &  \multicolumn{3}{c}{fMDCM (see $^b$)}\\
\hline \hline
\end{tabular}
$^a$\url{https://github.com/MMunibas/DES2/blob/main/M2/source/rkhs_SCN_rRz.csv}
$^b$\url{https://github.com/MMunibas/DES2/blob/main/M2/source/scn_fluc.dcm}
\end{table}

\begin{table}
\caption{Bonded and non-bonded parameters for Model {\bf M3} with
  TIP4P water model.  Nonbonded LJ parameters for SCN$^-$ are adopted
  from cluster interaction energy fit without scaling of
  $r_\mathrm{min}$.}
\label{sitab_params_m3}
\begin{tabular}{c|ccc}
\hline\hline
\textbf{Residues} & \multicolumn{3}{c}{Parameters} \\
\hline \hline
\textbf{Acetamide} & \multicolumn{3}{c}{CGenFF\cite{cgenff}} \\
\hline \hline
\textbf{Water} & \multicolumn{3}{c}{TIP4P\cite{TIP3P-Jorgensen-1983}} \\
\hline \hline
\textbf{K$^+$} & \multicolumn{3}{c}{} \\
\hline
Nonbonded\cite{bian:2013} & $\epsilon$ (kcal/mol)& $r_\mathrm{min}$ (\AA) & $q$ (e)\\
K$^+$ & $0.1004$ & $3.7378$ & $+1$ \\
\hline \hline 
\textbf{SCN$^-$} & \multicolumn{3}{c}{} \\
\hline
Bonded & \multicolumn{3}{c}{RKHS (see $^a$)} \\
\hline 
Nonbonded & $\epsilon$ (kcal/mol)& $r_\mathrm{min}$ (\AA) & $q$ (e) \\
S & $0.1118$ & $5.3984$ & $-$ \\
C & $0.0001$ & $3.1786$ & $-$ \\
N & $0.0037$ & $5.0388$ & $-$ \\
\hline 
Electrostatic Model &  \multicolumn{3}{c}{fMDCM (see $^b$)}\\
\hline \hline
\end{tabular}
$^a$\url{https://github.com/MMunibas/DES2/blob/main/M3/source/rkhs_SCN_rRz.csv}
$^b$\url{https://github.com/MMunibas/DES2/blob/main/M3/source/scn_fluc.dcm}
\end{table}

\clearpage

\begin{figure}
\centering
\includegraphics[width=0.90\textwidth]{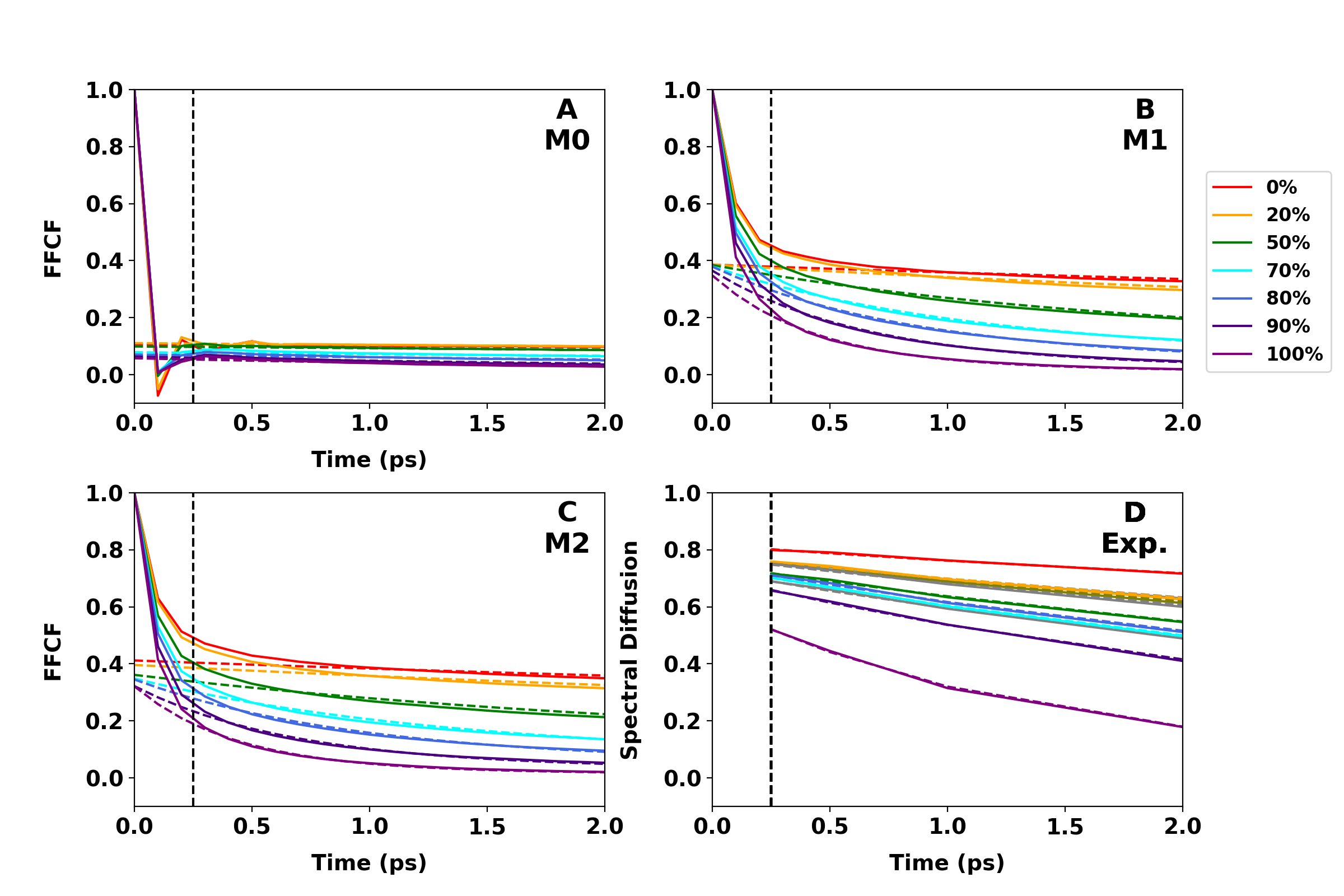}
\caption{Short time scale section ($< 2$\,ps) of the (A-C)
  Frequency-frequency correlation function (FFCF, solid lines) of the
  INM frequencies $\nu_3$ of SCN$^-$ anion from simulation with
  different force fields for different mixing ratios of acetamide and
  water. (D) Experimentally measured spectral diffusion. A
  bi-exponential function (dashed lines) is fit to each computed FFCF
  (A-C) and to the spectral diffusion (D).  The vertical dashed
    line marks the lower time boundary (0.25\,ps) to which experiments
    are sensitive.}
\label{sifig_t2ps_ffcf}
\end{figure}

\begin{figure}
\centering
\includegraphics[width=0.90\textwidth]{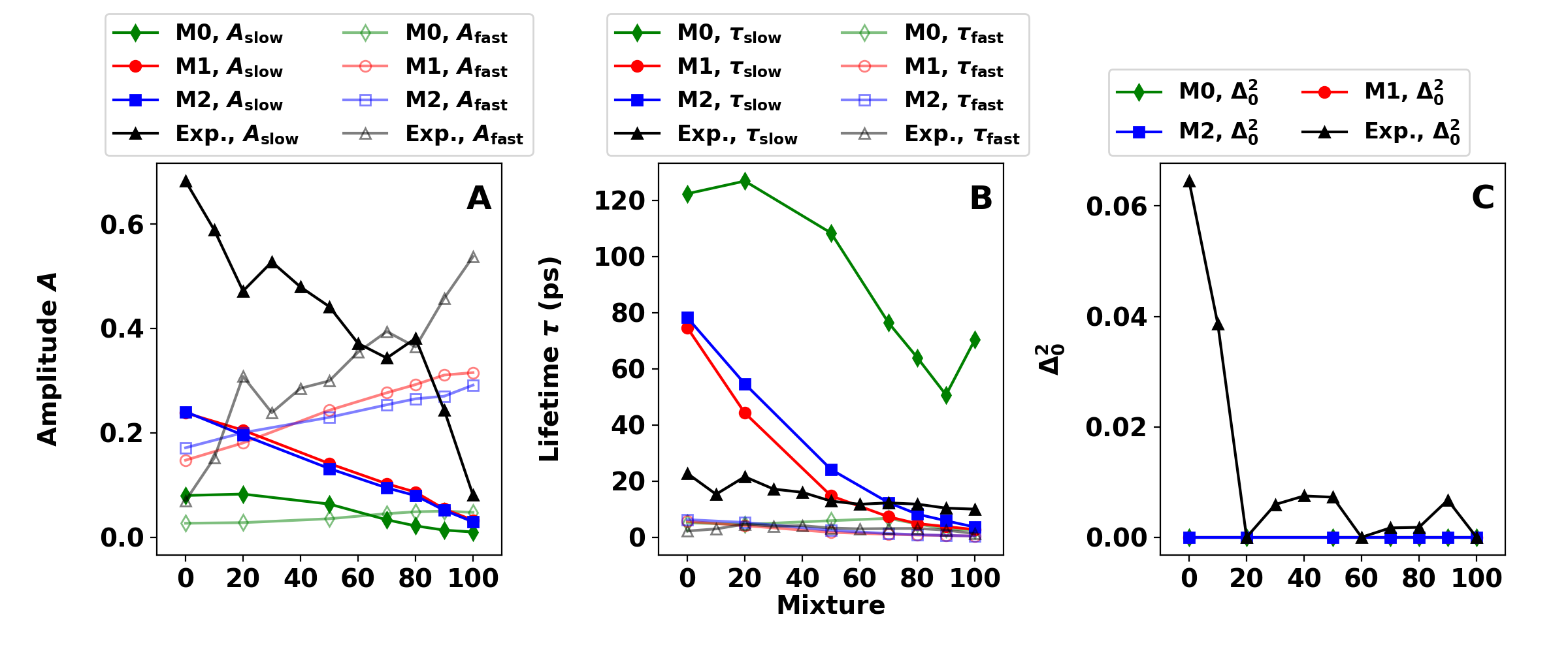}
\caption{Panels A to C: Optimized amplitudes, $(\tau_{\rm
    fast},\tau_{\rm slow})$ and $\Delta_0^2$ of the bi-exponential
  fits to the computed frequency-frequency correlation functions. The
  vibrational mode considered was $\nu_3$ of SCN$^-$ and frequencies
  were determined from instantaneous normal modes. Simulations were
  carried out using models {\bf M0} to {\bf M2}.}
\label{sifig_ffcf_fit}
\end{figure}

\begin{table}
\caption{
  Fitted lifetimes in ps to
  the ({\bf M0, M1, M2}) computed FFCFs and ({\bf Exp.})
  experimentally measured spectral diffusion from KSCN in
  water(W)/acetamide(ACM) mixtures for given W:ACM mixing ratios.}
\label{sitab_ffcf_fit}
\begin{tabular}{c||cc||cc||cc||cc}
\hline\hline
  & \multicolumn{2}{c||}{\bf M0} & \multicolumn{2}{c||}{\bf M1} & \multicolumn{2}{c||}{\bf M2} &\multicolumn{2}{c}{\bf Exp.} \\ \hline
W:ACM & 
  \multicolumn{1}{c}{$\tau_{\rm fast}$}   & \multicolumn{1}{c||}{$\tau_{\rm slow}$} &
  \multicolumn{1}{c}{$\tau_{\rm fast}$}   & \multicolumn{1}{c||}{$\tau_{\rm slow}$} &
  \multicolumn{1}{c}{$\tau_{\rm fast}$}   & \multicolumn{1}{c||}{$\tau_{\rm slow}$} &
  \multicolumn{1}{c}{$\tau_{\rm fast}$}   & \multicolumn{1}{c}{$\tau_{\rm slow}$} \\
\hline \hline
0 & 5.15 & 122.36 & 5.65 & 74.63 & 6.28 & 78.21 & 2.20 & 22.74 \\
10 & - & - & - & - & - & - & 2.97 & 15.30 \\
20 & 4.57 & 126.87 & 4.13 & 44.27 & 5.26 & 54.64 & 4.66 & 21.55 \\
30 & - & - & - & - & - & - & 3.92 & 17.16 \\
40 & - & - & - & - & - & - & 4.14 & 16.01 \\
50 & 5.90 & 108.37 & 1.75 & 14.76 & 2.47 & 24.14 & 3.28 & 12.87 \\
60 & - & - & - & - & - & - & 2.99 & 11.76 \\
70 & 6.73 & 76.39 & 1.06 & 7.28 & 1.31 & 12.33 & 3.10 & 12.24 \\
80 & 4.73 & 63.98 & 0.80 & 4.87 & 0.90 & 8.23 & 3.14 & 11.80 \\
90 & 2.99 & 50.69 & 0.62 & 3.81 & 0.65 & 5.87 & 2.54 & 10.37 \\
100 & 2.56 & 70.31 & 0.43 & 2.73 & 0.42 & 3.64 & 1.29 & 10.01 \\
\hline
\hline
\end{tabular}
\end{table}

\clearpage

\begin{figure}
\centering
\includegraphics[width=0.60\textwidth]{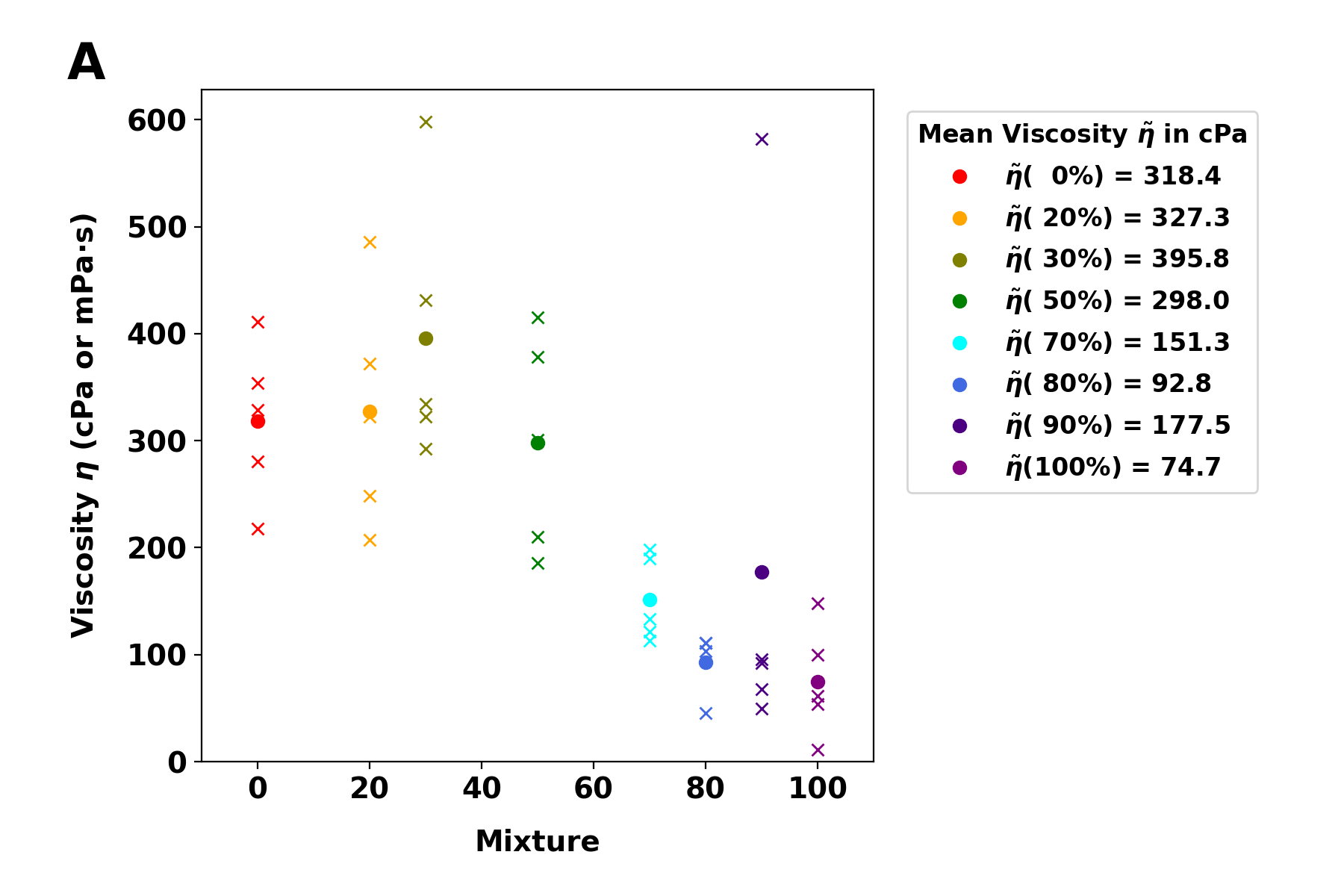}
\includegraphics[width=0.40\textwidth]{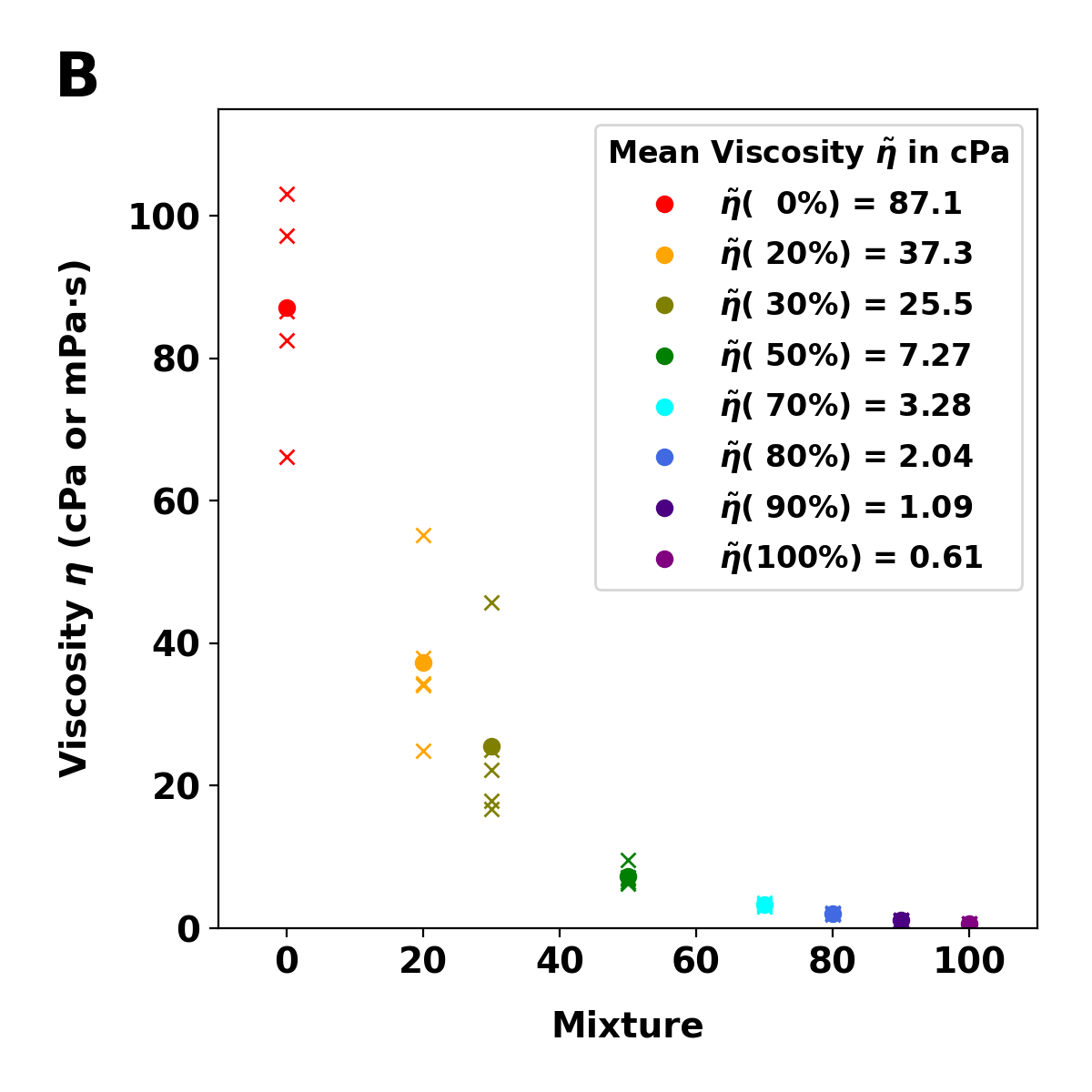}
\includegraphics[width=0.40\textwidth]{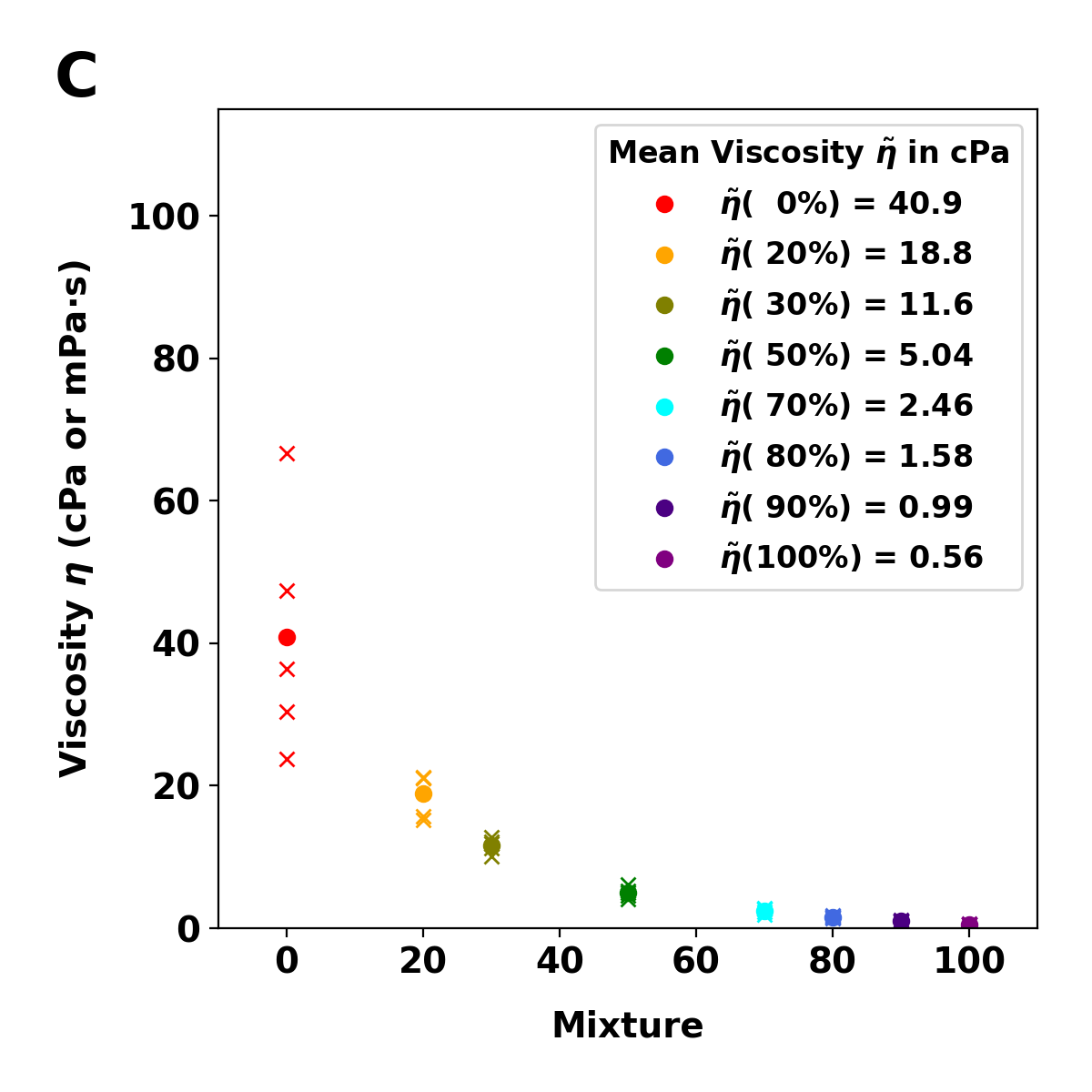}
\caption{Computed viscosity from the Green-Kubo relation (see main
  text) using the stress tensor correlation function from simulations
  using models (A) {\bf M0}, (B) {\bf M1} and (C) {\bf M2} for
  different acetamide/water mixtures. The results were obtained from 5
  individual runs of 5\,ns $NVT$ simulations each per mixture,
  restarting from the last state of the respective $NpT$
  simulation. The viscosity estimation per run are shown as colored
  crosses and their mean value as a colored circle. It is important to
  stress that the simulation length (5 ns) is not sufficient for
  convergence, in particular for low water content.}
\label{sifig_viscosity}
\end{figure}

\clearpage
\bibliography{refs}